\documentclass[a4paper,11pt]{article}
\pdfoutput=1 

\usepackage{jcappub} 

\usepackage[T1]{fontenc} 

\usepackage{amsmath}
\usepackage{amssymb}
\usepackage{amsfonts}
\usepackage{mathabx}
\usepackage{graphicx}  
\usepackage{xspace}			
\usepackage{wasysym}
\usepackage{float}
\usepackage{color}

\usepackage{mathtools}

\DeclarePairedDelimiter\floor{\lfloor}{\rfloor}

\definecolor{nicered}{rgb}{0.7,0.1,0.1}
\definecolor{Zpurple}{RGB}{119, 50, 168}

\graphicspath{{figs/}}			
\numberwithin{equation}{section}    

\newcommand{\acro}[1]{\textsc{\MakeLowercase{#1}}\xspace}

\title{Constraining Time Dependent Dark Matter Signals from the Sun}

\author[a,b,1]{Mohammadreza Zakeri,\note{Corresponding author.}}
\author[a,c,d,e]{Yu-Feng Zhou,}

\affiliation[a]{Institute of Theoretical Physics, Chinese Academy of Sciences,\\ ZhongGuanCun East Street 55, Beijing 100190, P. R. China}
\affiliation[b]{Department of Physics and Astronomy, University of Kentucky,\\ 505 Rose St., Lexington, KY 40506, USA}

\affiliation[c]{School of  Physics, University of Chinese Academy of Sciences,\\ Yuquan Road 19, Beijing 100049, China}

\affiliation[d]{School of Fundamental Physics and Mathematical Sciences, Hangzhou Institute for Advanced Study, UCAS\\ XiangShan Zhinong 1, Hangzhou 310024, China}

\affiliation[e]{International Centre for Theoretical Physics Asia-Pacific,\\ Beijing/Hangzhou, China}

\emailAdd{m.zakeri@uky.edu}
\emailAdd{yfzhou@itp.ac.cn}

\abstract{
  Dark matter (DM) particles captured by the Sun can produce high-energy electrons outside the Sun through annihilating into meta-stable mediators. The corresponding cosmic-ray electron signals observed by the space-based experiments will be time dependent due to the orbital motion of the space-based detectors. The shape of this time dependence is predictable given the orbital information of the detectors. Since the high-energy CR electron (with energy $E>100$ GeV) fluxes are expected to be constant in time, non-observation of such time variation can be used to place upper limits on the DM annihilation cross section. We analyze the time dependence of dark matter cosmic-ray signals in three space-based experiments: AMS-02, DAMPE and CALET. Under the assumption that no time dependent signal is observed, we derive the 95\% C.L. exclusion limits on the signal strength from the current data. We map our limits onto the parameter space of the dark photon model and find that the constraints are comparable with that derived from the supernova SN1987A.
}

\begin{document}
\maketitle
\flushbottom

\section{Introduction}

The origin, acceleration and propagation of cosmic rays (CRs) in the Milky Way have been the subject of cosmic ray measurements by detectors such as AMS, DAMPE, CALET, and Fermi-LAT. Supernova remnants are considered~\cite{Baade259} as the main sources for Galactic CRs. It's also broadly accepted~\cite{Fermi-LAT:2013iui} that the CRs are accelerated by the diffusive Fermi acceleration mechanism~\cite{Fermi:1949ee} as the supernova shock wave moves through the interstellar medium\footnote{For a critical review of this standard paradigm for the origin of the Galactic CRs check~\cite{Gabici:2019jvz}, and for an example of the calculation of the spectra for HE protons and nuclei see~\cite{Ptuskin:2010}.}. One of the main sources of errors in analyses of CRs comes from the uncertainties in the propagation models. Therefore, constraining the propagation parameters using our observations will reduce this uncertainty. For example, it was shown~\cite{Jin:2014ica} that the boron to carbon ratio ($B/C$) data combined with the proton flux measurements from AMS-02 can be used to determine the propagation parameters. There are also ambiguities in modeling the acceleration and propagation effects in the heliosphere especially for low energy CRs. Studying the solar modulation effects on the CRs can advance our understanding of the cosmic ray problems. As an example, the Voyager-1 spacecraft measurements of low energy CRs and their time dependence~\cite{Wang:2019xtu, Zhu:2018jbk} are very significant in analyzing the local interstellar spectra and solar modulation effects on CRs. We can also investigate the  propagation of different particle species in the Milky Way and the charge-sign dependency of the solar modulation effects by studying the differences of modulation effects among electrons, positrons, and nuclei~\cite{Zhu:2020koq}. On the high-energy (HE) side of the spectrum, since the Sun is not expected to be a source of HE CRs, probing their solar modulation can aid us in constraining possible new physics sources such as DM. Even though these DM sources have no explicit time-dependence, the CR signals observed by the space-based detectors will have a time dependence induced by their orbital motion. The background HE CRs are expected to be isotropic and constant in time. Therefore, a non-observation of time dependencies in the observed signals can be used to constrain the time-dependent portion of the signal generated by new physics (DM). In this work we examine the bounds on the strength of solar HE CRs modulations using the data from AMS, DAMPE, CALET and assume that no time-dependent signals have been observed. The model independent bounds presented here, can be mapped onto the specific parameter spaces of various DM models.

A diverse set of experiments and observations have been searching for dark matter (DM) signals but none has detected any signals yet. The hunt for DM continues in three fronts: direct detection, collider searches, and indirect detection. Indirect detection of DM could lead to information about its annihilation cross section and mass. Indirect searches are mainly carried out by observations of gamma rays and other CRs, and can explore higher energies, longer decay lengths, and weaker interactions compared to other searches~\cite{Leane:2020liq}. The main obstacle in the indirect search analyses is the lack of a good understanding of the backgrounds, which introduces large systematic errors. It is therefore clear that we should aim to maximize the signal to background ratio. One of the ways to accomplish this is to look at a scenario which has a distinct signal shape from that of the background spectrum. This is usually done by comparing the signal and background energy spectra. In the absence of such DM signals, we can still constrain DM models given an understanding of the astrophysical backgrounds~\cite{Jin:2013nta}. DM searches with charged CRs are capable of reaching extremely high energies which are far above our terrestrial collider energies. However, since the CR propagation is not well understood large systematic uncertainties are induced in these searches. Among charged CRs, HE electrons and positrons play a crucial role in our understanding of nearby sources of CRs. This is due to the fact that they lose their energy rapidly as they travel through the Galaxy, e.g., via synchrotron radiation and inverse Compton scattering off of the interstellar radiation field~\cite{1970ApJ162L181S}.  For instance, cosmic ray $e^{\pm}$ reaching the Earth with energies about $1$ TeV must have originated from nearby sources that are less than $0.75$ kpc away~\cite{Ackermann:2010ip}. The Sun is not believed to be a source of HE $e^{\pm}$ CRs, and those $e^{\pm}$ produced in cosmic ray interactions with the surface of the Sun are expected to be trapped due its high magnetic field. This alleviates the background estimation issue. In this work, we focus on the Sun as a possible nearby source for high-energy $e^{\pm}$ signals which are generated by the DM accumulated in the Sun. We explore the distinction between the signal and background as a function of both time and energy.

Astrophysical objects can act as a gravitational potential well for the DM particles around them. If DM particles have interactions with the Standard Model particles other than gravity, they will scatter off of matter inside these objects. Those scatterings which lead to DM particles' final velocities below that of the escape velocity in the object, will result in the capture of DM particles~\cite{Press:1985ug, Gould:1987ir}. Dark matter particles lose their energy after multiple scatterings and accumulate at the center of the objects. If the captured DM self-annihilates, the capture rate will balance the annihilation rate resulting in an equilibrium. The formalism for the DM capture in the large astrophysical bodies has been analyzed long time ago~\cite{Freese:1985qw,Press:1985ug,Silk:1985ax, Krauss:1985aaa,Griest:1986yu,Gaisser:1986ha,Gould:1987ju, Gould:1987ir,Gould:1987ww,Gould:1991hx}. The early works studied the annihilation of the accumulated DM in the Sun to neutrinos (e.g. for explaining the solar neutrino problem). The neutrinos produced by the DM annihilation can escape the Sun and be detected at Earth, while any other Standard Model products will not be able to escape the Sun. Alternatively, DM may annihilate into new long-lived mediators which can escape the Sun and decay into SM particles outside the Sun~\cite{Batell:2009zp,Schuster:2009au,Schuster:2009fc,Meade:2009mu,Feng:2016ijc, Bell:2021pyy}. It is also shown that Jupiter can be a viable DM target especially for sub-GeV DM owing to its cooler core temperature compared to the Sun~\cite{Leane:2021tjj}. These scenarios are realized in the context of secluded DM models.

Dark sectors secluded from the Standard Model and equipped with a metastable mediator (between the visible and dark sectors), can evade the laboratory-based experimental bounds by having a small coupling between the mediator and the visible sector~\cite{Pospelov:2007mp}. One signature of such models is the DM annihilation into the mediator and the subsequent decay of the mediator into the SM particles. In general we expect this annihilation to occur at the Galactic center and dwarf spheroidal satellites galaxies (dSphs). Therefore, any model generating signals in the Galactic center should also be consistent with the lack of such signals from dSphs. A consistent example of this is studied in~\cite{Ding:2021zzg} as a possible explanation to the observations of an excess of positron flux by PAMELA~\cite{PAMELA:2013vxg}, Fermi-LAT~\cite{Fermi-LAT:2011baq} and AMS-02~\cite{AMS:2019rhg}.
In the specific case of dark photon models, the dark photon can decay into a pair of electron-positron which could be detected by the detectors around the Earth. This type of signal from the Sun at the Alpha Magnetic Spectrometer (AMS-02) was studied in~\cite{Feng:2016ijc}. 

We propose a new method to optimize the search for this type of secluded DM models using the modulations of high-energy CR signals from the Sun that are induced by the orbital motion of space-based detectors. Accordingly, we evaluate the exposure to the Sun as a function of time in three experiments, and use this exposure to find the number of signal events in their detectors. We exploit this time dependence and study its behavior by varying the time-binning schemes and comparing the limits on the signal strength of the model. Our constraints are highly model-independent and can be easily translated into specific DM models, e.g., dark photon model. We present our bounds in terms of the annihilation rate ($\Gamma_{\rm ann}$) which is the most direct and model-independent way of constraining DM signals from the Sun. We then translate these bounds into the spin-dependent dark matter nucleon cross section.

We conclude the introduction by summarizing the structure of this paper. In the next section~(\ref{sec:signal}) we talk about the exposure of the detector to the Sun which is detailed in appendix~\ref{app:sunexp}. In section~\ref{sec:indep}, we find the model-independent limits on the signal strength using the data from the three detectors AMS-02, DAMPE, and CALET. We will then discuss how cosmic ray signals could be generated in a generic class of models and present our limits in their parameter space in section~\ref{sec:generic}. In the following section (\ref{sec:dark}) we present our limits on the parameter spaces of the dark photon models. This is followed by a conclusion and the appendices which include the details of the calculations for the position of the Sun (appendix~\ref{app:sunpos}) and the detectors' exposure to it (appendix~\ref{app:sunexp}).

\section{Signals at Satellites}
\label{sec:signal}

\begin{figure}
  \centering
 \includegraphics[width=0.84\textwidth]{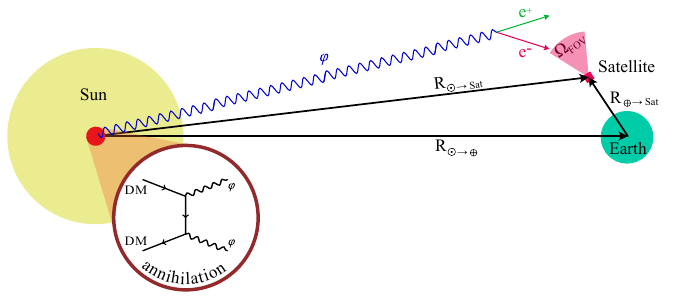}
 \caption{\label{fig:ess}Schematic diagram depicting the CR production process originating from the Sun and their subsequent detection in the detectors around the Earth.}
\end{figure}

The number of signals detected by the detectors is in general a function of the model (as detailed in section~\ref{sec:generic}), and the detector parameters such as the exposure to the Sun ($\xi$). This exposure depends on the effective acceptance of the detector($\mathcal{A}$), exposure time intervals ($T$), and the orbit of the detector between time interval (bin) $\Delta T = [t_1, t_2]$, and energy bin $[E_1, E_2]$. The exposure time intervals are a function of energy and the orbit of the detector which include the geomagnetic effects. We will approximate these effects by including an overall factor ($r_t < 1$) in the exposure~\cite{Aguilar:2018ons}, such that the effective exposure time is $T = r_t\times\Delta T$. The effective acceptance can also be approximated by its value at the characteristic energy $\tilde{E}$ (i.e., spectrally weighted mean energy in each bin):
\begin{eqnarray}
  \label{eq:exp}
  \xi(E_1, E_2, t_1, t_2) = \frac{\mathcal{A} (\tilde{E})}{\Omega_{\rm FOV}} \times r_t \times \int_{t_1}^{t_2} dt\, \left( \frac{1 \rm AU}{|\vec{R}_{\Sun \to \rm Sat}(t)|}\right)^2\bigg[-\cos(i(t))\bigg]_{i\in \Omega_{\rm FOV}},
\end{eqnarray} 
in which $\vec{R}_{\Sun \to \rm Sat}(t)$ is the vector pointing from the Sun to the detector (Fig.~\ref{fig:ess}), $i(t)$ is the angle between $\vec{R}_{\Sun \to \rm Sat}(t)$ and the normal to the detector's surface ($\hat{n}$), which is constrained to be within the field-of-view (FOV) of the detector ($\Omega_{\rm FOV}$). The ratio of AU to the distance of the Sun to the detector is a correction factor for the time-dependence of the distance to the Sun. The effect of the Earth's shadow is also implicitly included in Eq.~\ref{eq:exp}. Note that the exact FOV can itself depend on the energies and type of the incident CRs. In this work we only use a simple conical approximation of FOV. A more detailed calculation of the exposure of detectors to the Sun is given in appendix~\ref{app:sunexp}.

The orbit of the detector around the Earth and Earth around the Sun induces a specific time dependence to the signal. Therefore, we propose using the time-binned data to constrain the signal strength. In the absence of time analyses of HE CRs in experiments, we can assume that all of the results so far show no time dependence. In other words, we divide each energy bin ($i$-index) into smaller time bins ($j$-index), with equal time intervals ($\delta t$) and number of events $n_{ij} \pm \delta n_{ij}$, such that

\begin{eqnarray}
  n_{ij} =& (\delta t / T) \times N_i,\quad \delta n_{ij} =& \sqrt{\delta t / T} \times \delta N_i,
\end{eqnarray}
in which $T$ is the total experiment time. We then model the number of observed events in each bin as
\begin{eqnarray}
  n_{ij} = \mu\times \frac{\xi_{ij}}{4\pi \left(1\, {\rm AU}\right)^2} + B_{ij}, \label{eq:sig:def}
\end{eqnarray}
in which $\mu$ is the signal strength which is a global fitting parameter, $\xi_{ij}$ is the shape (exposure) of the signal events which is fixed by the time-binning. The overall scale of the signal is fixed by $\mu$, which absorbs all the model-dependencies. $B$ is the number of background events which itself is taken to be a fitting parameter. Since we are assuming no time dependence is observed, the result of fitting $\mu$, and $B_{ij}$ is trivially given by $\mu = 0$, and $B_{ij} = n_{ij}$. The 95\%~\acro{C.L.} on $\mu$ is then extracted from the upper error ($\delta\mu_{\rm up}$) of the fitted signal strength $\mu$.

The next question is how do we define the time bins? Given a periodic signal, the best upper limit on the signal strength is extracted if the time-bins are shorter than the period of the signal. At the same time, the time bins have to be chosen wider at higher energies since we are constrained by the number of observed events. In order to ensure that we have enough events in all of the bins, we can divide the lower energy bins into more bins (shorter time periods), with longer bin periods for the higher energy bins. 

We use likelihood ratio test $\chi^2_{\lambda}$ to fit the signal strength $\mu$, and find its $95\%~\acro{C.L.}$ We therefore minimize 

\begin{eqnarray}
\chi^2_{\lambda} = -2\ln \prod_{ij} \frac{L(n_{ij} \,;\, \mu\times S_{ij} + B_{ij}) }{ L(n_{ij} \,;\, n_{ij}) },  \label{eq:likli}
\end{eqnarray}
in which the likelihood $L$ is given by
\begin{eqnarray}
  L = \prod_{ij} \frac{e^{-(\mu\times S_{ij} + B_{ij})}}{n_{ij}!}\left(\mu\times S_{ij} + B_{ij} \right)^{n_{ij}},
\end{eqnarray}
and $95\%~\acro{C.L.}$ limits can be found by setting $\Delta(-2\ln L) = F^{-1}_{\chi^2_n}(0.95)$, in which $n$ is the number of fit parameters. We now discuss the relevant details of each of the three detectors under consideration.  

\textbf{AMS-02:} Precision measurements of the time and energy dependence of the cosmic ray $e^{\pm}$ flux from May 2011 to May 2017 have been performed by AMS-02 for the energy range from $1$ to $50$ GeV~\cite{Aguilar:2018ons}. The events from every Bartel (27 days) were combined into one time-bin. A time-dependent modulation happens when interstellar CRs enter the heliosphere~\cite{Potgieter:2013pdj}. Among various processes involved in the solar modulation, particle drift creates a dependence on the particle charge~\cite{Potgieter:2014pka}. Given that electron and positron have opposite charge but the same mass, the measurement presented in~\cite{Aguilar:2018ons} leads to crucial information about the charge-sign dependent solar modulation effects. From Fig.~1 of~\cite{Aguilar:2018ons}, we can see that the time-variation of the signal is notable for energies below $10$ GeV, and vanishes for $30$ GeV $ \lessapprox E $. In this work, we will be focusing on much higher energies. For a more recent analysis of the cosmic ray spectra and their solar modulation at AMS-02 check Ref.~\cite{Zhu:2020koq}.

The total duration of the AMS-02 data~\cite{Aguilar:2019ksn} is 2370 days. AMS-02 which was originally planned to have a three-year life in space without maintenance and then wind down, has been under maintenance to extend its life-time. For the future projection of the bounds, here we have assumed that the AMS-02 life-time can be extended through 2028\footnote{The extension of ISS operation through 2028 has been introduced in H.R.5666~\cite{hr5666}.}.

We focus on the last $13$ bins of the AMS-02 data~\cite{Aguilar:2019ksn} covering the energy range $[98.1, 1400]$ GeV. We assume a time-binning of each of them according to 
\begin{eqnarray}
  \label{eq:ams:tbin}
  \delta t = \{ 10, 10, 10, 10, 15, 15, 15, 15, 15, 30, 45, 80, 240 \} \qquad \rm Days,
\end{eqnarray}
in the order of increasing energy. In choosing the time intervals we are constrained by two main factors. The time interval should be long enough to ensure enough statistics. Furthermore, for our assumption about the active data taking time being a constant factor times the total time, we need enough time so that the orbital anomalies average out over many orbit cycles. Therefore, we don't take time-bins shorter than $10$ days. Further improvements could possibly be made by including the real-time data from these detectors in our program which is beyond the scope of this work. For the exposure details see~\ref{app:sunexp:ams}.

\textbf{CALET:} The total duration of the CALET data~\cite{Adriani:2018ktz} is 780 days. We consider the last $21$ bins which covers energies $[94.9, 4754.7]$ GeV. The following time-bins are assumed:
\begin{eqnarray}
  \delta t = \{ 18, 18, 18, 18, 18, 18, 18, 18, 18, 20, 20, 20, 30, 36, 36, 36, 60, 60, 90, 300, 360 \} \,\, \rm Days.
\end{eqnarray}
For more details on the exposure of CALET to the Sun see~\ref{app:sunexp:calet}.

\textbf{DAMPE:} The total duration of the DAMPE data~\cite{Ambrosi:2017wek} is 530 days. There is a hint of a narrow excess at energy $\sim 1.4$ TeV. It is shown~\cite{Huang:2017egk} that this peak can appear in two scenarios. In one of these scenarios, the peak is generated from the ``spectrum broadening" of continuous sources with delta-function like injection spectrum, e.g. DM annihilation into $e^+e^-$ in the subhalos. It has been shown that these sources must be nearby $\sim 0.3$ kpc, and the DM annihilation cross sections are close to the thermal value. We consider the last $28$ bins covering an energy range of $[95.5, 4570.9]$ GeV. These bins are further time-binned according to
\begin{eqnarray}
  \delta t = \{ &10,& 10, 10, 10, 10, 10, 10, 10 ,10, 10, 10, 18, 18, 18, 18,\nonumber\\
                &30,& 30, 90, 100, 100, 265, 265, 265, 265, 530, 530, 530, 530 \} \quad \rm Days.
\end{eqnarray}
For more details on the exposure of DAMPE to the Sun see~\ref{app:sunexp:dampe}.

We would like to mention the other very different time-dependent dark matter signal which is caused by the rotation of the Earth around the Sun\footnote{Check~\cite{Froborg:2020tdh} for an overview of  annual modulation measurements.}, i.e., yearly modulations. The modulating signal in that case is from the DM particles scattering off of targets in direct search experiments such as DAMA/LIBRA~\cite{DAMA:2008jlt}. 

\section{Model Independent Analysis}
\label{sec:indep}
In this section we present the results of minimizing Eq.~\ref{eq:likli} in a model independent manner. The minimization of Eq.~\ref{eq:likli} is done with respect to the signal strength $\mu$ and the background parameters $B_{ij}$ using {\tt Minuit} package~\cite{James:1975dr}. We then find their corresponding $95\%$ error bars using {\tt Minos} error analysis. This limit on the signal strength ($\mu_{95}$) can be translated in terms of the specific model parameters at hand. We will demonstrate this further in sections~\ref{sec:generic} and~\ref{sec:dark}, after presenting the model independent bounds on $\mu$ from each of the three detectors in this section. 

\begin{figure}
  \centering
 \includegraphics[width=1\textwidth]{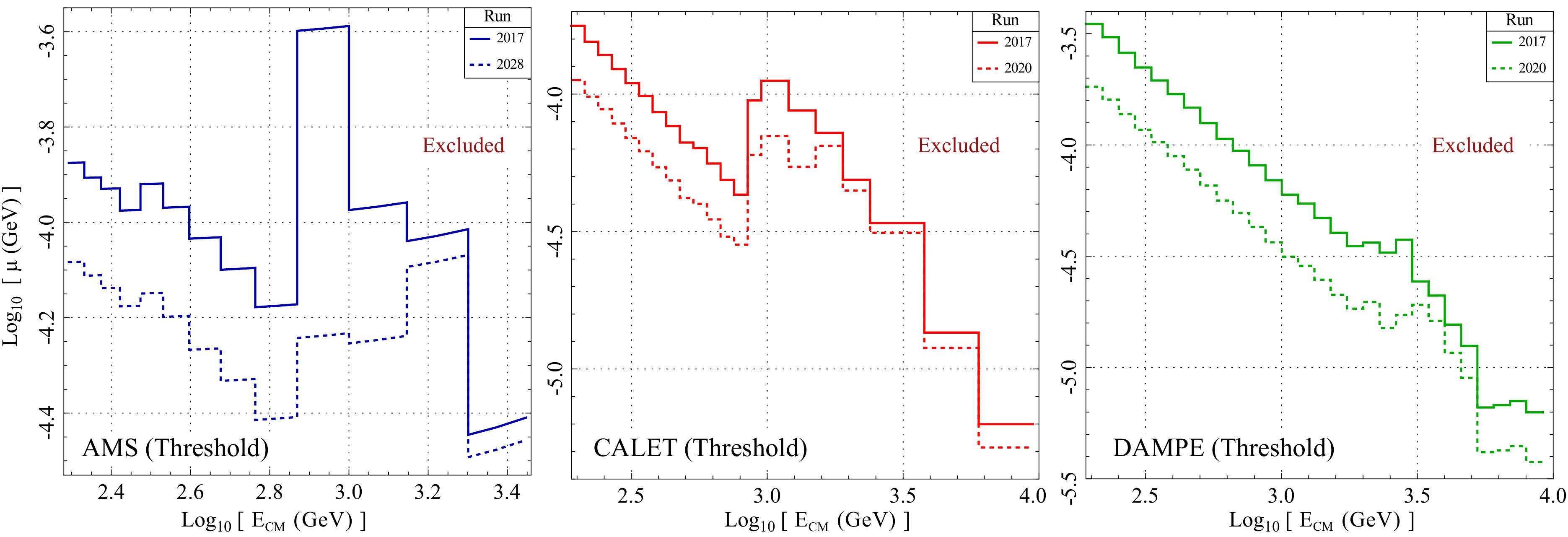}
 \hfill
 \includegraphics[width=1\textwidth]{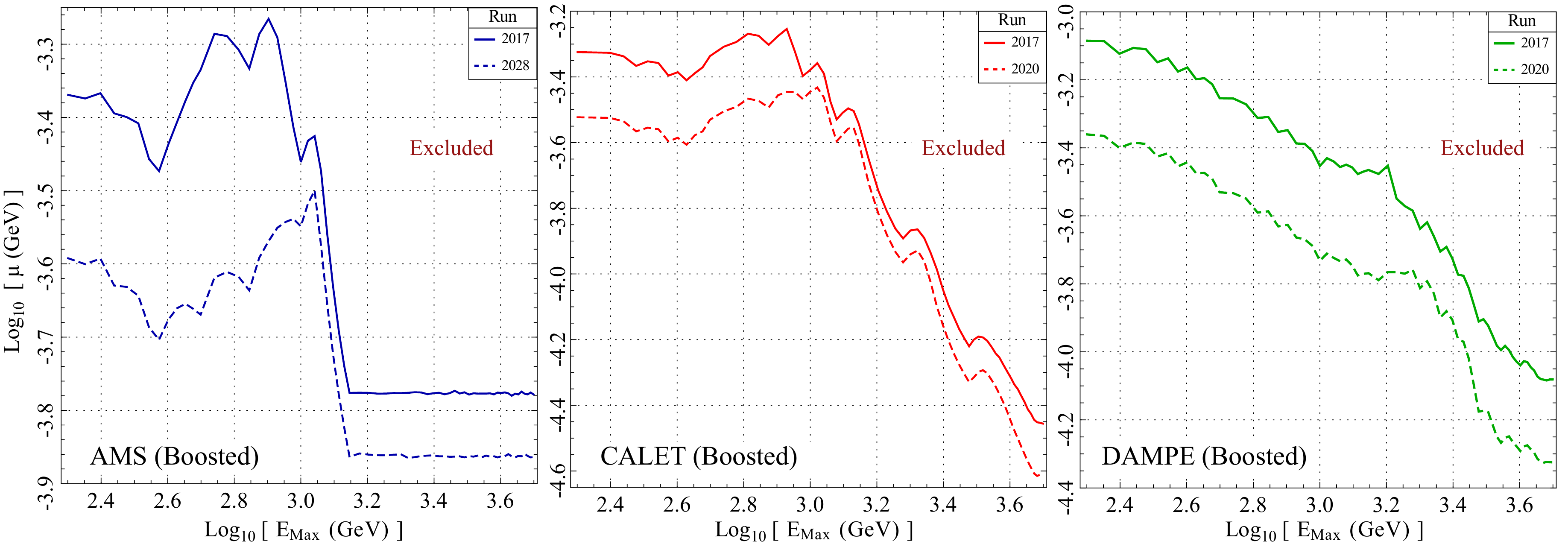}
 \caption{\label{fig:mu}Limits on the signal strength ($\mu$) at $95\%$~\acro{C.L.} as a function of the CR energy ($E$) using the current (solid) data and future projections (dashed) in the threshold (top) and boosted (bottom) scenarios. The areas above the curves are excluded.}
\end{figure}

We are interested in box-shaped (``Boosted'') and delta-like signals (``Threshold''). In the threshold case, the signal spectrum is sharp and only in a single energy bin\footnote{For example if the threshold signal is resulting from the 2-body decay of a mediator with energy $E_{\rm CM}$, the decay products will each have an energy equal to $E_{\rm CM}/2$.}($E = E_{\rm CM}/2$).  The limits from each of the detectors on the signal strength ($\mu$) in the threshold case are plotted in Fig.~\ref{fig:mu}. The current bounds are shown as solid blue curves, whereas the future projections are shown with dashed blue curves.

In the boosted case the signal spectrum is uniformly distributed between some $E_{\rm Min}$ to $E_{\rm Max}$. Therefore, multiple energy bins will contain the signal and among them the bins with higher energies (lower background) will be influencing the constraints the most. The resulting limits on the signal strength ($\mu$) in the boosted case are plotted in Fig.~\ref{fig:mu}. We would like to clarify that even though the boosted case is similar to the one considered in~\cite{Cuoco:2019mlb, Mazziotta:2020zte}, the threshold case is different from the ``delta-like" signals in~\cite{Cuoco:2019mlb, Mazziotta:2020zte}. 

For AMS and in the boosted case, the limit becomes constant for DM masses above $\sim 1.4$ TeV. This is because for $M_{\rm DM} > 1.4$ TeV, the signal will cover all the energy bins, and increasing the DM mass doesn't change the statistics anymore. For the CALET future projections (red-dashed), we have assumed a total mission life of $5$ years. We can see that the constraints are stronger than AMS-02, especially at higher energies.

Note that the model-independent bounds on the signal strength ($\mu$) are independent of the specific magnetic effects that would be incorporated in the mediator decay probability ($P_{\rm dec}$) as in~\cite{Feng:2016ijc}.
\section{Generic Models}
\label{sec:generic}
In this section we will consider generic models in which the annihilation rate of the DM in the Sun is used as an independent parameter. In this way, we are agnostic about the specific and model-dependent mechanisms involving the DM capture, and annihilation in the Sun. We only demand that a generic model implements the following processes:
\begin{enumerate}
  \item Dark matter ($X$) captured in the Sun annihilates ($\Gamma_{\rm ann}$) to mediators ($\varphi$)
  \item Mediators ($\varphi$) travel radially outward from the Sun
  \item Mediators ($\varphi$) then decay to $e^{\pm}$
\end{enumerate} 
These $e^{\pm}$ may then be detected by detectors around the Earth. The condition for the mediator ($\varphi$) to escape the Sun with no significant interactions can be used to set an upper limit on the cross section between the mediator and the Sun's constituents. Since the captured dark matter is concentrated at the center of the Sun within a radius $r_X \ll R_{\Sun}$, the resulting mediators would be emanated from the center in radial directions. The average number of mediator interactions on their way out of the Sun is given by
\begin{equation}
    \textrm{\# of scatterings} = \sum_i \int_0^{R_{\Sun}} n_i(r) \sigma_i dr,
\end{equation}
in which the sum runs over the various species that interact with the mediator, $n_i(r)$ is their number density at radius $r$, and $\sigma_i$ is their scattering cross section which in general depends on the energy of the scattering target (i.e., temperature) and thus may implicitly depend on the radial position ($r$). Given the DM mass ranges that we are considering, the mediator energies are much larger than the thermal energy of the targets and we can ignore this radial dependence. We examine the constituents at nucleon level and expand the summation above in terms of scattering cross sections off of nucleons and electrons. The contribution from each term must be smaller than $1$ in order for the mediator to freely escape from the Sun's interior. Specifically for electron and proton targets we must have
\begin{equation}
    \sigma^{-1}_{p, e} \gtrapprox \int_0^{R_{\Sun}} \rho(r) \left[\frac{X(r)}{m_{\rm H}} +  \frac{2Y(r)}{m_{\rm He}}\right] dr,
\end{equation}
in which $\rho(r)$ is the mass density of the Sun, $\{ m_{\rm H}$, $X(r)\}$ and $\{m_{\rm He}$, $Y(r)\}$ are the hydrogen and helium atomic masses and mass fractions respectively. A similar result for the neutron scatterings can be obtained by removing the hydrogen contributions. We input the values for $\rho(r), X(r), Y(r)$ from the BP2004 solar model~\cite{Bahcall:2004fg}, and find $\sigma_{p,e, n} \lessapprox 10^{-36}\, {\rm cm}^2$. Given that $R_{\Sun} \approx 2.32$ light-second, and $1$ AU $\approx 499$ light-second, in order for the mediator to decay between the Sun and Earth we should have
\begin{equation}
    2.3\, (s) <  \left[\sqrt{1 - \left(m_{\varphi}/m_X\right)^2}\right]\, \tau < 499\, (s),
\end{equation}
in which $\tau = (m_X/m_{\varphi}) \tau_{0}$ is the mediator's lifetime in the Sun's rest frame, and $\tau_{0}$ is its rest-frame lifetime. We are interested in the limit $m_{\varphi} \ll m_X$, for which the mediators are highly boosted and travel at speeds close to $c$. If we take $M_X \approx \mathcal{O}({\rm TeV})$, and $m_{\varphi}\approx \mathcal{O}({\rm MeV})$, we see that $\tau_0$ must be between $2.3\, \mu s$ and $0.5$ ms.

Given that the evaporation of DM particles can be neglected for DM masses above a few GeV~\cite{Griest:1986yu, Gaisser:1986ha}, the number of DM particles in the Sun is governed by

\begin{equation}
  \frac{dN_{\rm DM}}{dt} = \Gamma_{\rm cap} - C_{\rm ann}\, N^2_{\rm DM},
\end{equation}
in which the capture rate is given by $\Gamma_{\rm cap}$, and the annihilation rate is encoded in $C_{\rm ann}$. If an equilibrium is established, i.e., $\dot{N}_{\rm DM} = 0$, we get
\begin{equation}
  \Gamma_{\rm ann} = \frac{1}{2} C_{\rm ann}\, N^2_{\rm DM} = \frac{1}{2} \Gamma_{\rm cap},
\end{equation}
in which we have included a factor of $1/2$ for the fact that each annihilation involves two DM particles. The total number of potential signal events produced is then given by 

\begin{eqnarray}
  N_S = N_S^0 \, \rm Br(\varphi \to e^{+} e^-) \times P_{\rm dec}, \label{eq:ns}
\end{eqnarray}
in which $Br(\varphi \to e^{+} e^-)$ is the branching ratio of the mediator to $e^{+} e^-$, and $N_S^0$ is the number of mediators produced which is given by  
\begin{eqnarray}
  N_S^0 = n_{\varphi} \times\Gamma_{\rm ann} \times \frac{\xi_{\Sun}}{4\pi\left(1\, {\rm AU}\right)^2}. \label{eq:ns0}
\end{eqnarray} 
Here, $n_{\varphi}$ is the number of $\varphi$'s produced per DM annihilation, $\xi_{\Sun}$ is the detector's exposure to the Sun, and $P_{\rm dec}$ is the probability for the mediator to decay within a specific distance and energy range. We can write it in terms of a probability density for the mediator decay products to be produced at a distance $r_d$ and energy $E = {E_1, E_2}$
\begin{eqnarray}
  P_{\rm dec} = n_e\times \int dr_d dE\, \frac{dP_{\rm dec}}{dr_d dE} = n_e\times \int dr_d dE\, \frac{\exp(-r_d/L)}{L} \times f(E),
\end{eqnarray} 
in which $f(E)$ is the energy spectrum of the decay products which depends on the angle ($\theta_{\rm CM}$) between the mediator boost direction and one of the final states in the center-of-mass (CM) frame. We assume that the decay products are evenly distributed in $\cos\theta_{\rm CM}$, and we will show below how this determines the form of $f(E)$ in the threshold and boosted cases. The factor $n_{e}$ is the number of detectable final states ($2$ for $\varphi \to e^+ e^-$ at DAMPE). Note that the two-particle (or more) events are a small fraction of the total= number of single-particle events, due to the angular separation between the decay products $\sim\mathcal{O}(\rm mrad)$, which yields distances much larger than the effective size of the detectors $\sim \mathcal{O}(\rm m)$~\cite{Feng:2015hja}. Comparing equations~\ref{eq:ns} and \ref{eq:ns0} with \ref{eq:sig:def} we see that the signal strength is identified as 

\begin{eqnarray}
  \mu \coloneqq n_{\varphi} \times\Gamma_{\rm ann}\,\times \rm Br(\varphi \to e^{+} e^-) \times P_{\rm dec},
\end{eqnarray} 
such that limits on $\mu$ can be converted into contours in the parameter space of our model, e.g., in terms of $\Gamma_{\rm ann}$ and $L_{\rm dec}$.

We focus our attention on two-body decays of $\varphi$ in this work, but the results can be extended to more complicated spectra. Specifically, we look at two kinematical regimes 

\begin{itemize}
  \item \textbf{Boosted:} $ 2m_e \ll m_{\varphi} \ll m_X$
  \item \textbf{Threshold:} $ 2m_e \approx m_{\varphi} \ll m_X$
\end{itemize}
We note that since dark matter is much heavier than the mediator the dark matter annihilation is Sommerfeld enhanced by a factor $S\gg1$. This enhancement shortens the equilibrium time-scale ($\tau \sim 1/\sqrt{S\, C_{\rm ann}}$) inside the Sun, and it enters the annihilation rate as
\begin{equation}
    \Gamma_{\rm ann} = \frac{\Gamma_{\rm cap}}{2} \tanh^2 \left(\frac{t}{\tau}\right).
\end{equation}
Since for the ranges of parameter space that are of interest, the equilibrium must have already been achieved in the Sun, i.e., $t \gg \tau$, the effects of Sommerfeld enhancements are negligible. Nevertheless, we have included the Sommerfeld enhancement effect in our calculations for completeness. We will be focusing on energies greater than $\sim 100$ GeV, which would not include the low-energy time dependent signals due to solar activity~\cite{Aguilar:2018ons}. The boosted case generates a rectangular energy spectrum and the decay products can be detected in multiple energy bins (for $E \in [E_{\rm cut}, m_X]$), whereas the threshold case produces a delta function spectrum $\delta(E - m_X/2)$ with a sharp signal in the energy bin containing $E = m_X/2$. In order to see this, note that since $m_{\varphi} \ll m_X$ \& $E_{X} = E_{\varphi} \approx m_X$ in the Sun's reference frame, $\varphi$'s are boosted by $\gamma = m_{X}/m_{\varphi}$ from their CM reference frame, in which $E'_{e^{\pm}} = m_{\varphi} / 2$. In the threshold case, $2m_e \approx m_{\varphi}$ so $e_{\pm}$ are produced at rest with respect to $\varphi$, i.e., they are boosted by the same factor $\gamma$. Therefore, their energy in the Sun's reference frame is given by $E_{e} = \gamma \times m_{\varphi} / 2 = m_X/2$. 
In the boosted case, the energies of the decay products in the Sun's frame is given by $(1\pm \cos \theta_{\rm CM})m_X/2$. Here, $\theta_{\rm CM}$ is the angle between the mediator boost direction and one of the decay products, and it's uniformly distributed. Therefore, the energies of the decay products are uniformly distributed in $[0, m_X]$. For more details about the decay spectra, such as their angular and velocity distributions we refer the reader to the appendix in Ref.~\cite{Feng:2015hja}. In summary, we have:
\begin{align}
  P_{\rm dec}^T =& \exp(-R_{\Sun}/L) - \exp(-1 {\rm AU}/L) \quad& \rm (Threshold)\\
  P_{\rm dec}^B =& \left[\exp(-R_{\Sun}/L) - \exp(-1 {\rm AU}/L) \right] \times \frac{E_{\rm max}-E_{\rm min}}{m_X} \quad& \rm (Boosted) \label{eq:p:boost}
\end{align}
in which $E_{\rm max, \rm min}$ are defined as
\begin{align}
  E_{\rm min} = {\rm max} (E_1, E_{\rm cut}), \qquad  E_{\rm max} = {\rm min} (E_2, m_X),
\end{align}
with $E_{1, 2}$ are the energy limits of the bin under consideration.

Note that the solar magnetic field from the Sun and Earth can deflect the trajectory of the mediator decay products. Consequently, their trajectory will not point exactly back to the Sun when they are detected at the Earth. These effects are especially important for search proposals involving angular cuts~\cite{Feng:2016ijc}, in which simplified models of the solar magnetic field effects are utilized. An angular cut may then enhance the search for signals pointing back to the Sun. Since the current data that we use doesn't have any directional features, we will not impose any angular cuts, and therefore the magnetic field effects can be ignored.

In the threshold case, in which we expect a concentration of signals in a single energy bin ($E_i$), the mediator mass must be within a few keV range from $2m_e$. Otherwise, the boost factor will spread the signal to other neighboring bins. Nevertheless, we show the results for mediator masses up to $\sim 5$ MeV recognizing that the results are more accurate for $m_{\varphi} \approx 2m_e$. We will present the results of minimization of Eq.~\eqref{eq:likli} in the generic models for the three detectors under consideration. In the boosted case we have chosen a representative set of DM masses from $200$ GeV to $5$ TeV. In the threshold case, the contours correspond to the limits from each of the energy bins. The DM mass labels in this case represent the corresponding energy bin containing $E = E_{\rm CM}/2 = m_{\rm DM}/2$. For example, the AMS-02 exclusion contour for $m_{\rm DM} = 1.2$ TeV also applies to other DM masses within $[1.0,  1.4]$ TeV range.

\subsection{Threshold Case}
\begin{figure}
  \centering
 \includegraphics[width=0.95\textwidth]{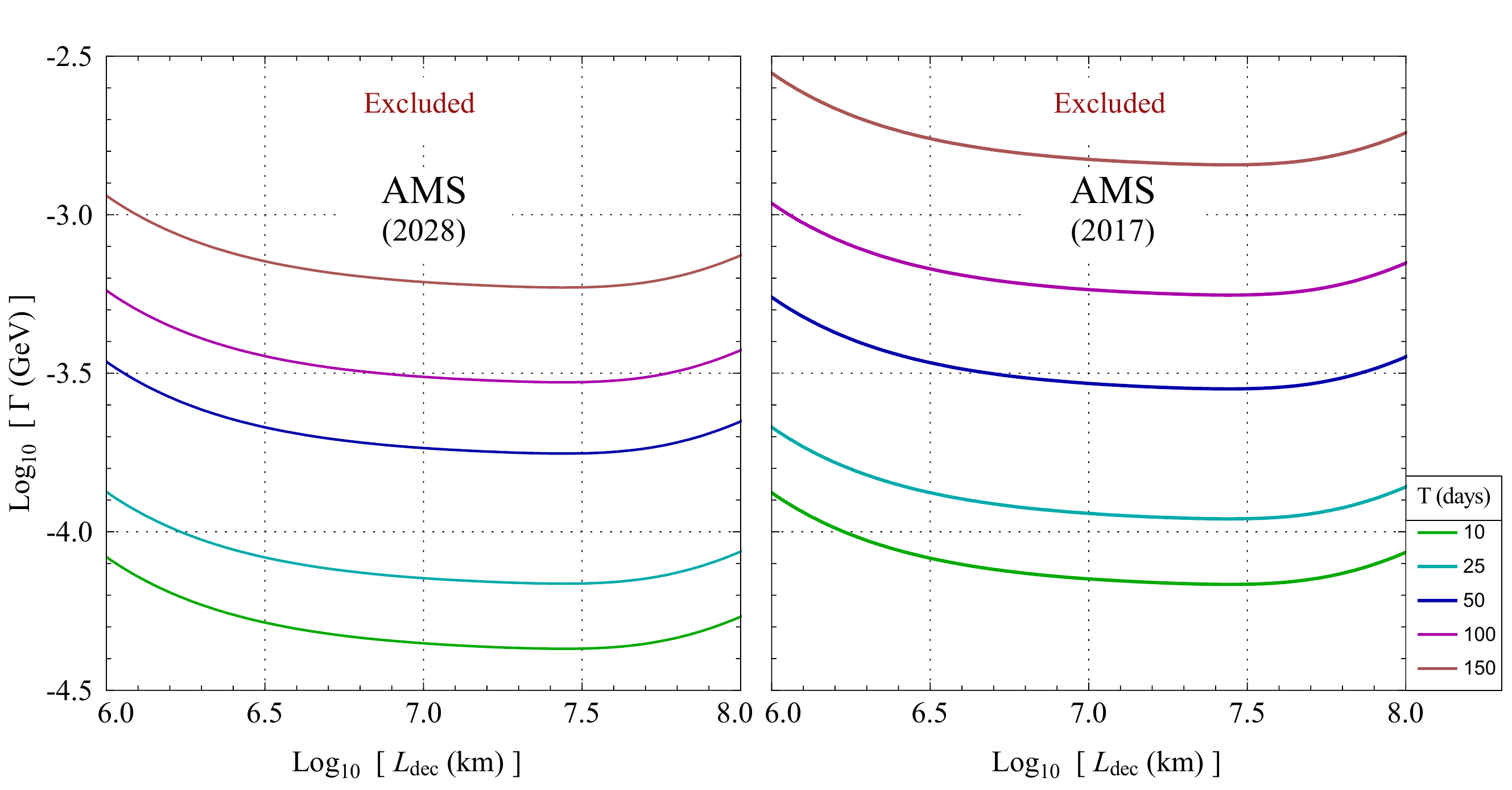}
 \caption{\label{fig:ams:gen:tbin}Limits on the annihilation rate ($\Gamma_{\rm ann}$) at $95\%$~\acro{C.L.} as a function of the decay length ($L_{\rm dec}$) for various time binning using the $E \in [ 98.1 - 107.3 ]$ GeV bin from AMS-02 data, in the threshold scenario. \textbf{Left:} Future projections. \textbf{Right:} Current bounds.}
\end{figure}

In order to demonstrate how time-binning enhances the limits, we plot the $95\%$ contours for one of the energy bins with $E \in [ 98.1 - 107.3 ]$ GeV, corresponding to different time-bin periods $\{10, 25, 50, 100, 150 \}$ days, in the threshold scenario in Fig.~\ref{fig:ams:gen:tbin}. The area above each curve is excluded. We can see that shorter time periods yield better limits, specifically we can achieve a factor of $\sim 15$ improvement (see Fig.~\ref{fig:ams:gen:tbin}). In the boosted case, it's not straightforward to make this comparison as the signal covers many energy bins each with a different time-binning. The improvement will also depend on the DM mass. For heavier DM candidates, the higher energy bins are populated which have lower backgrounds. These low background bins determine the constraint on the signal strength. Due to a low count in these bins, the time binning will be limited to ensure enough statistics in each time-bin. As a result, we expect a better improvement for lighter DM candidates in the boosted case. 

\begin{figure}
  \centering
  \includegraphics[width=1\textwidth]{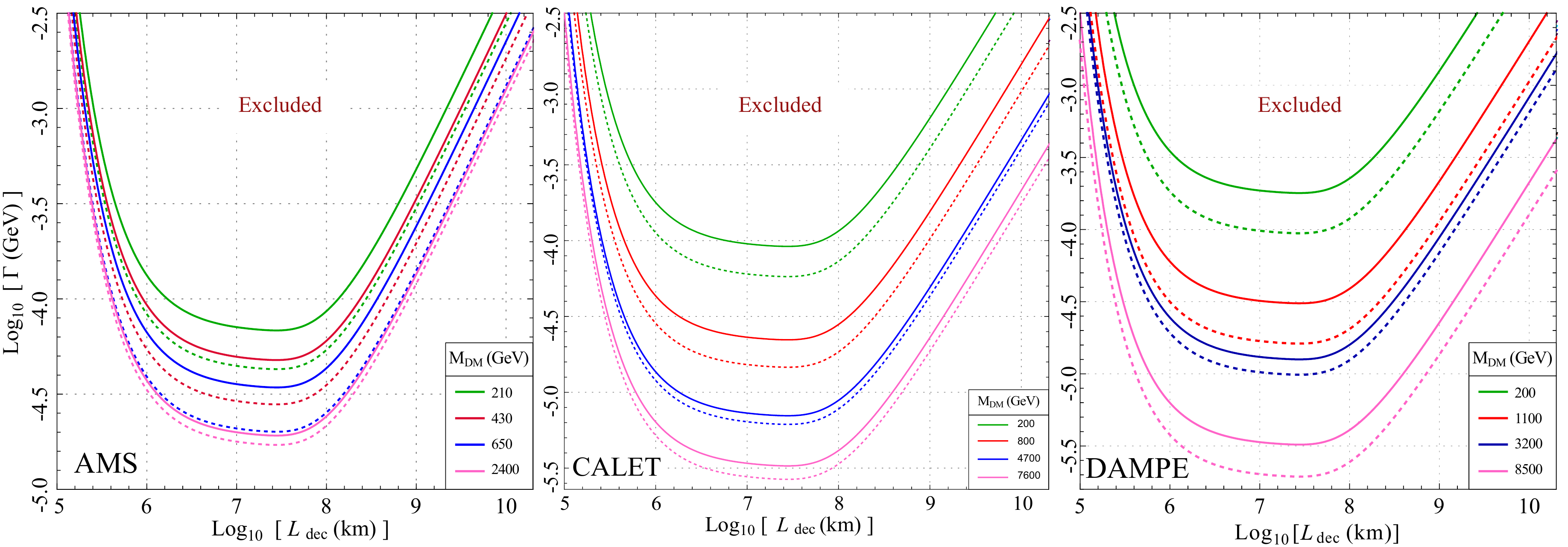}
  \includegraphics[width=1\textwidth]{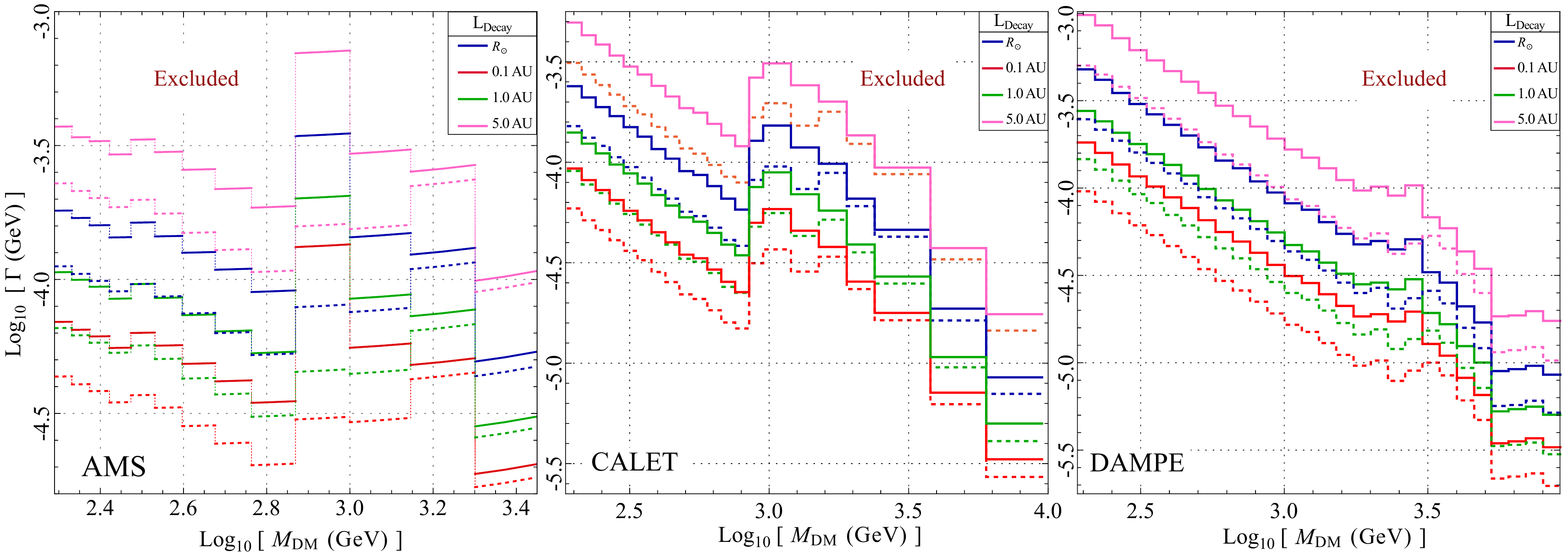}
 \caption{\label{fig:gen:thr}Limits on the annihilation rate ($\Gamma_{\rm ann}$) at $95\%$~\acro{C.L.} as a function of the   decay length (top) and the DM mass (bottom) using the current (solid) data and future projections (dashed) in the threshold scenario. The areas above the curves are excluded.}
\end{figure}

The resulting $95\%$ bounds in the threshold case are plotted in Figs.~\ref{fig:gen:thr}, in which the area above each curve is excluded. We see that for a  fixed time-bin period, the limit improves as we increase the DM masses which is due to a decrease in the background. However, in order to ensure enough statistics in higher energy bins, we are forced to make the time-bins longer for heavier DM masses which loosens the constraints on the signal strength. 

\begin{figure}
  \centering
  \includegraphics[width=1\textwidth]{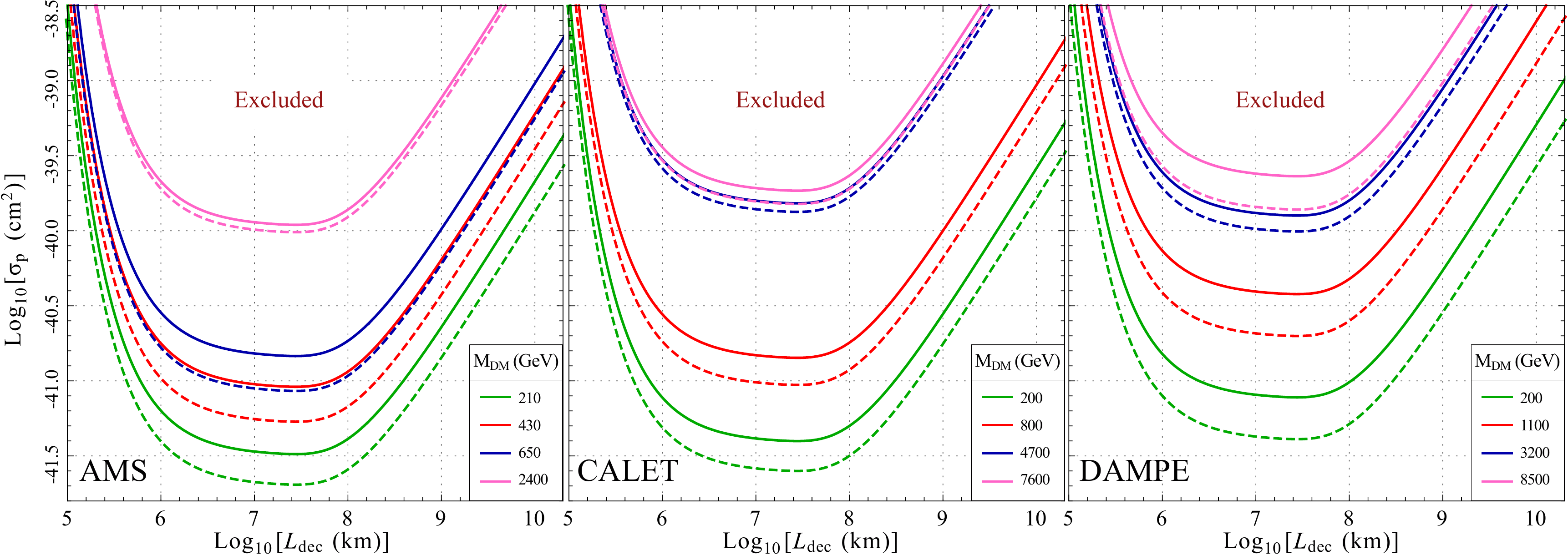}
  \includegraphics[width=1\textwidth]{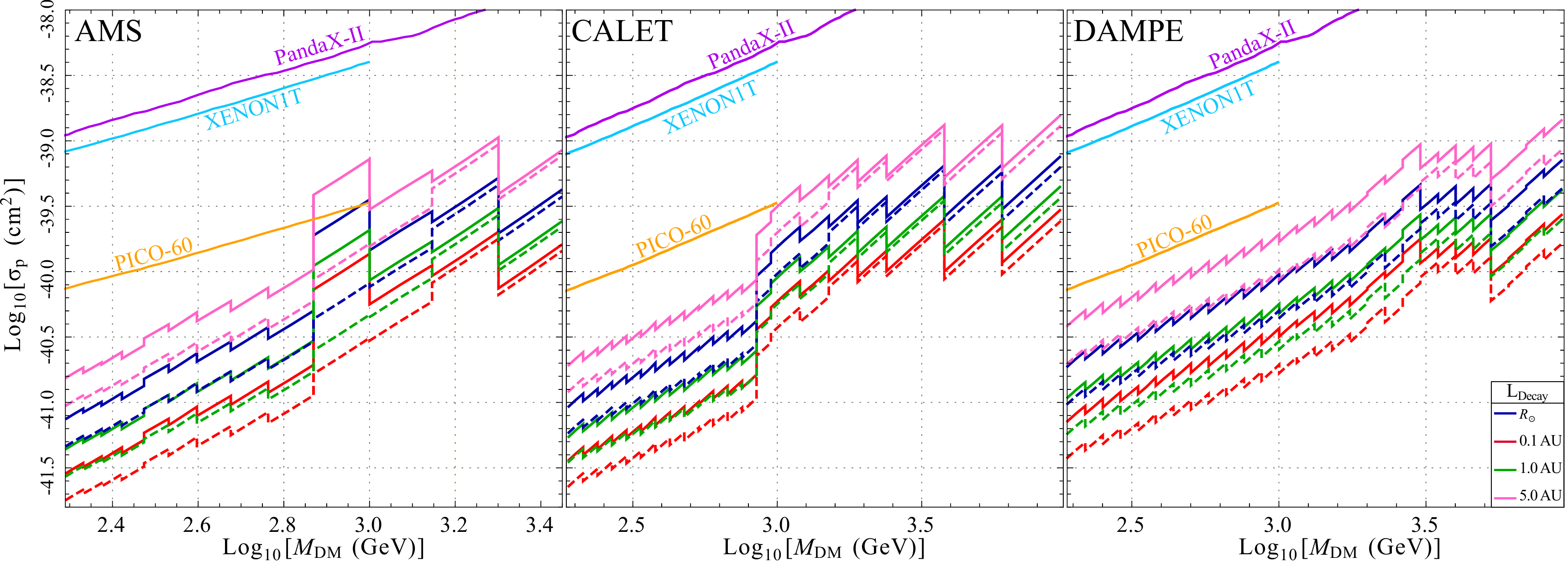}
 \caption{\label{fig:gen:thr:xsec}Limits on the spin-dependent DM-proton scattering cross section ($\sigma_{p}$) at $95\%$~\acro{C.L.} as a function of the decay length (top) and the DM mass (bottom) using the current (solid) data and future projections (dashed) in the threshold scenario. The areas above the curves are excluded. The direct detection constraints are from PandaX-II~\cite{PandaX-II:2018woa}, XENON1T~\cite{XENON:2019rxp}, and PICO-60~\cite{PICO:2019vsc}. }
\end{figure}

The annihilation rate ($\Gamma_{\rm ann}$) can be evaluated as a function of the DM mass. It would also depend on the DM density and velocity around the Sun, and the DM-nucleon cross section. Once the annihilation rate is calculated for a specific cross section, it can then be scaled for other cross sections. For instance, Ref.~\cite{Cuoco:2019mlb} used {\tt DARKSUSY}~\cite{Bringmann:2018lay, Gondolo:2004sc} with the canonical values for the DM around the Sun to evaluate the DM capture rate. The results are presented in Fig.~2 of~\cite{Cuoco:2019mlb} with a DM-nucleon scattering cross section of $\sigma = 10^{-40}$ ${\rm cm}^2$, which can be used to translate our results into both spin-dependent and spin-independent limits. There are strong bounds on dark matter models with spin-independent interactions from direct search experiments. In the case of spin-dependent (SD) interactions, bounds on $\sigma^{\rm SD}$ in the long-lived mediator models were set using the Fermi-LAT and HAWC data~\cite{Leane:2017vag, HAWC:2018szf}. We specifically focus on the SD scenario, in which the capture is mostly done by the hydrogen~\cite{Baratella:2013fya} in the Sun. Therefore, we present our limits on SD DM-proton scattering in Fig.~\ref{fig:gen:thr:xsec} (threshold) and Fig.~\ref{fig:gen:boo:xsec} (boosted). We have also included the direct detection constraints from PandaX-II~\cite{PandaX-II:2018woa}, XENON1T~\cite{XENON:2019rxp}, and PICO-60~\cite{PICO:2019vsc}. We can see that our study in general sets better constraints than the ones from these direct searches. In the boosted scenario, and for $M_{\rm DM} < 1$ TeV, the bound from PICO-60 becomes competitive but is still weaker than our best limits with the optimal decay lengths.

\subsection{Boosted Case} 
\begin{figure}
  \centering
  \includegraphics[width=1\textwidth]{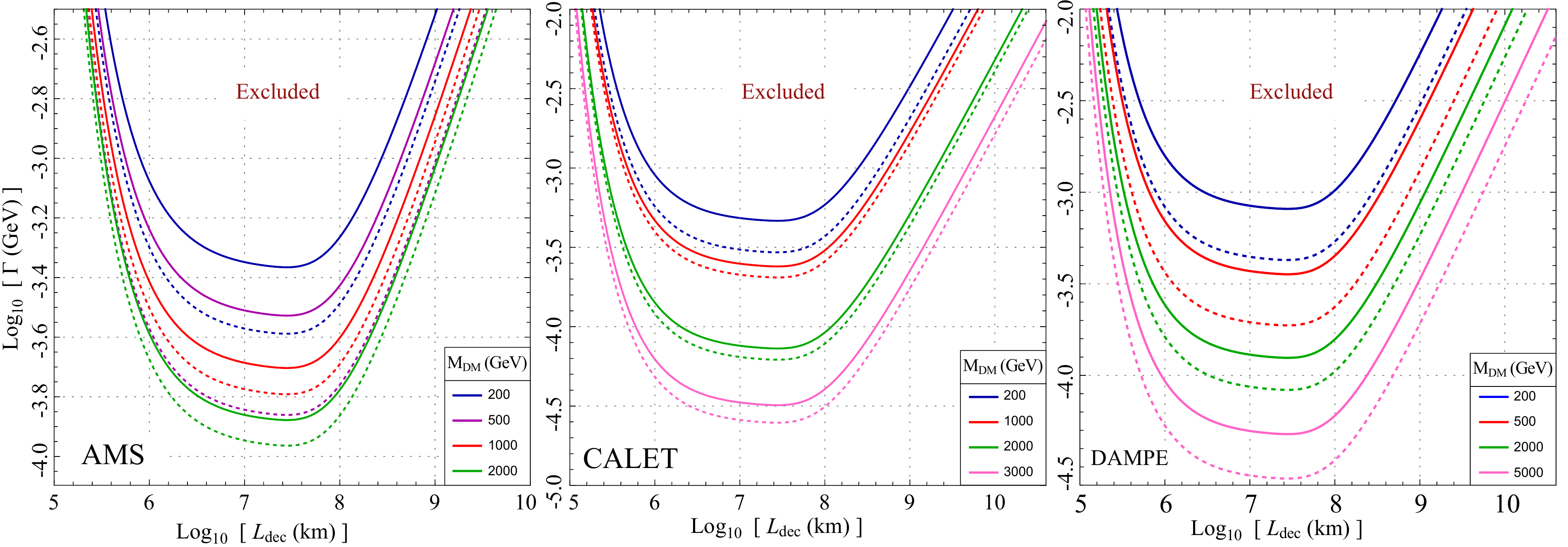}
  \includegraphics[width=1\textwidth]{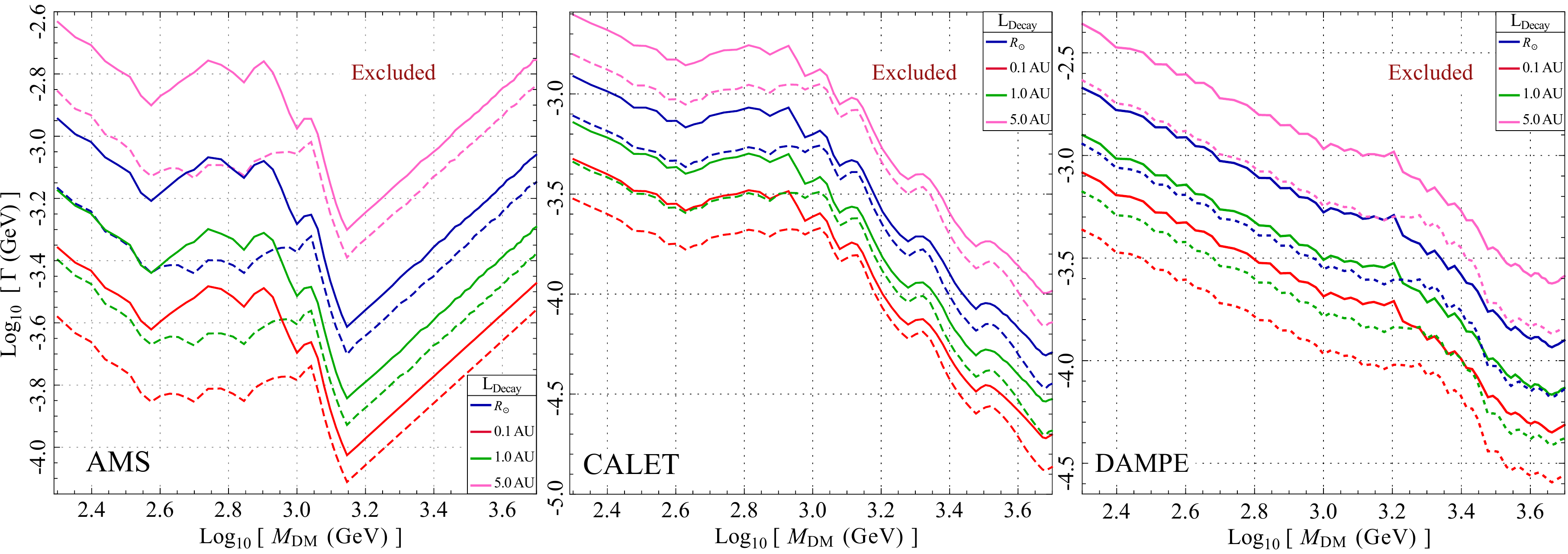}
 \caption{\label{fig:gen:boo}Limits on the annihilation rate ($\Gamma_{\rm ann}$) at $95\%$~\acro{C.L.} as a function of the decay length (top) and the DM mass (bottom) using the current (solid) data and future projections (dashed) in the boosted scenario. The areas above the curves are excluded.}
\end{figure}

\begin{figure}
  \centering
  \includegraphics[width=1\textwidth]{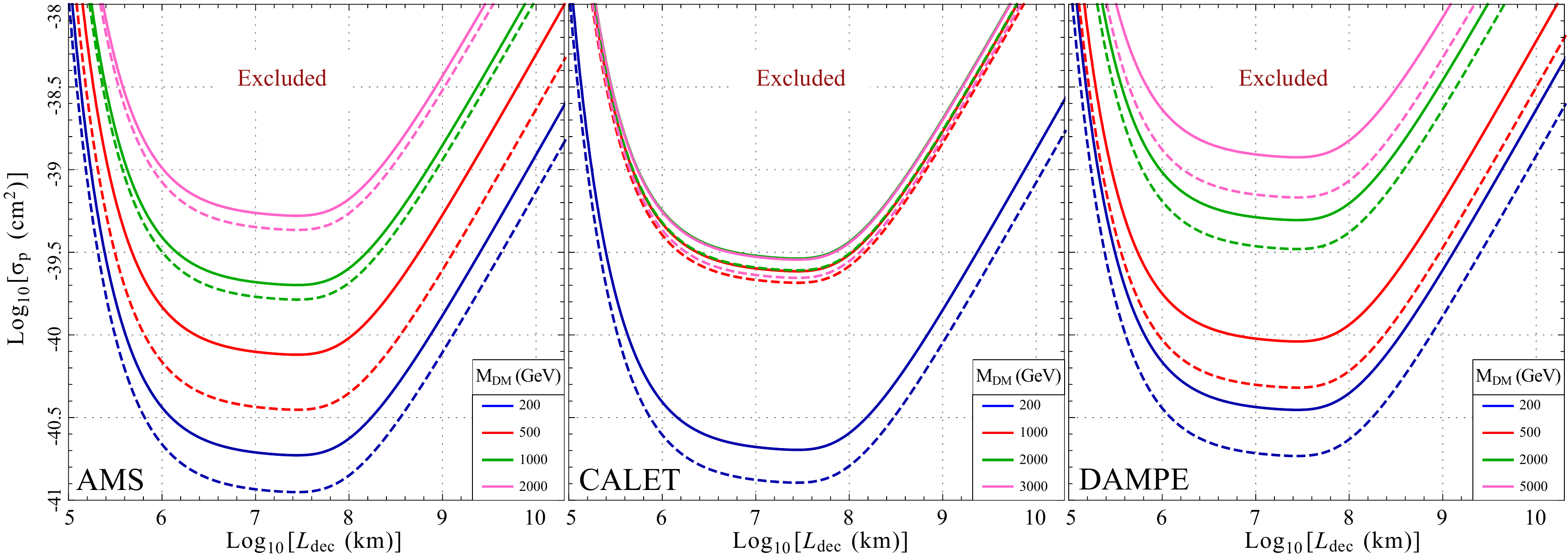}
  \includegraphics[width=1\textwidth]{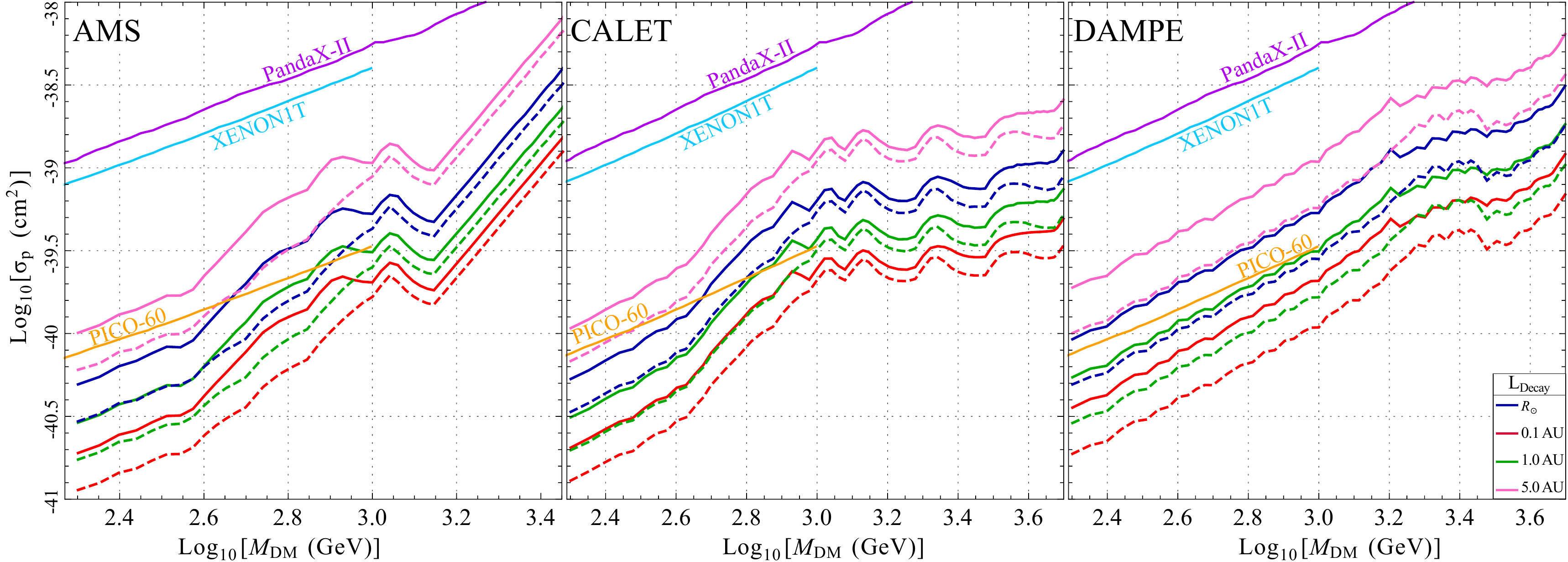}
 \caption{\label{fig:gen:boo:xsec}Limits on the spin-dependent DM-proton scattering cross section ($\sigma_{p}$) at $95\%$~\acro{C.L.} as a function of the decay length (top) and the DM mass (bottom) using the current (solid) data and future projections (dashed) in the boosted scenario. The areas above the curves are excluded. The direct detection constraints are from PandaX-II~\cite{PandaX-II:2018woa}, XENON1T~\cite{XENON:2019rxp}, and PICO-60~\cite{PICO:2019vsc}.}
\end{figure}

The constraints in the boosted scenario are plotted in Fig.~\ref{fig:gen:boo} and Fig.~\ref{fig:gen:boo:xsec}. In this case, and as we increase the DM mass the limits get better initially since we are adding the bins with lower counts. However, at the same time the signal strength which is normalized by the mass (Eq.~\ref{eq:p:boost}) will decrease. This is why the limits become weaker as we increase DM masses beyond $\sim 1.5$ TeV. 

One notable difference between the CALET and AMS-02 constraints, is the improvement of the constraints in the boosted case for CALET especially for heavier DM masses. This is because CALET has a higher event count compared to the AMS-02, and also its energy bins extend all the way to about $5$ TeV compared to $1.4$ TeV in AMS-02. For example, for $M_{\rm DM} = 5$ TeV, the boosted signal extends all the way to $5$ TeV, which is mostly covered by CALET, but the signals between $1.4 - 5 $ TeV  (i.e., $72\%$) will not be detected in AMS-02. We can see that the DAMPE bounds in the boosted case are weaker than the ones from CALET. This can be mainly attributed to the fact that the exposure of CALET to the Sun is more than the exposure of DAMPE, especially at higher energies. 

This work is complementary to the studies of similar models and their signals in Fermi-LAT~\cite{Mazziotta:2020zte}. Before we move on to the dark photon model, we would like to compare the results of this section to the ones found in the boosted case (``box-like'') from Fermi-LAT data~\cite{Mazziotta:2020zte} with a signal region defined as $30^{\degree}$ cone around the Sun. They have set limits on the DM-nucleon cross section, using a cone with aperture of $30^{\degree}$ pointing towards the Sun as the signal region. A ``box-like'' energy spectrum was considered in~\cite{Cuoco:2019mlb, Mazziotta:2020zte} which corresponds to the ``boosted" case in this work. Note that the ``delta-like'' spectrum in~\cite{Cuoco:2019mlb, Mazziotta:2020zte} corresponds to the capture via inelastic scattering and the subsequent annihilation directly to $e^+e^-$ pairs outside the Sun. Therefore, this ``delta-like'' signal is different from our ``threshold'' scenario which also produces a line-like signal. Multiple regions of interest (RoI) $[ 2^{\degree}, 5^{\degree}, 10^{\degree}, 30^{\degree}, 45^{\degree} ]$ have been assumed in~\cite{Cuoco:2019mlb}, with a signal region pointing towards the Sun and a control region in the anti-Sun direction. The angular cuts would in principle change the box-shaped spectrum, since $e^{\pm}$ with lower energies would be dispersed to outside the signal region. However, as long as the mediators are highly boosted, i.e., much lighter than the DM particles, the dispersion angle is small enough for the signal region to contain all the decay products.

The $95\%$~\acro{C.L.} on the elastic spin-independent DM-nucleon cross section for DM masses in $m_{\rm DM} \in [ 100, 1000 ]$ GeV range, for $L_{\rm dec} = 0.1$ AU is about $\sigma_{\rm SI} \sim (1-2)\times 10^{-45} \, {\rm cm}^2$. Using the fact that the capture rate is linear in $\sigma$, and its value as a function of the DM mass for $\sigma_{\rm SI} \sim 10^{-40} \, {\rm cm}^2$ in Fig.~2 of Ref.~\cite{Cuoco:2019mlb}, we get $\Gamma_{\rm cap} \sim  [3 \times 10^{19}, 5 \times 10^{18}] \, {\rm s}^{-1}$ or $\Gamma_{\rm cap} \sim  [2\times 10^{-5}, 3 \times 10^{-6}] \, {\rm GeV}$ for $m_{\rm DM} = 200$ GeV and $m_{\rm DM} = 1000$ GeV respectively. The equilibrium between capture and annihilation of DM particles in the Sun requires $\Gamma_{\rm ann} = 0.5 \Gamma_{\rm cap}$, we then get the Fermi-LAT limits of $\Gamma_{\rm ann} \sim  [10^{-5}, 1.5\times 10^{-6}] \, {\rm GeV}$ (for $m_{\rm DM} = 200$ GeV and $m_{\rm DM} = 1000$ GeV respectively) which is about $1$-$2$ orders of magnitude stronger than the limits we have obtained here for the same DM masses. For heavier DM masses, better limits on $\sigma_{\rm SI}$ are set in Ref.~\cite{Bell:2021pyy} using the HAWC observatory gamma ray data~\cite{HAWC:2018rpf}. Assuming that the probability of the mediator decay between the Sun and Earth is equal to $1$, limits are one order of magnitude better than this Fermi-LAT bound for $M_{\rm DM} \sim 1$ TeV. This assumption is not true in realistic models such as the dark photon model. In fact, we can see that the lower boundary of the constrained region in~\cite{Bell:2021pyy} on the dark photon parameter space lies within the SN1987A bounds and is around the same order of magnitude as our limits.

\section{Dark Photon Model}
\label{sec:dark}
In this section, we consider dark photon~\cite{Kobzarev:1966qya, Okun:1982xi} as a concrete example and translate the bounds from the previous section into limits on dark photon parameter space. We only present the main definitions for the dark photon model here, and the reader can check~\cite{Feng:2016ijc} for more details. The Lagrangian for photon-dark photon interaction is given by
\begin{eqnarray}
  \mathcal{L} = -\frac{1}{4} F^{\mu\nu}F_{\mu\nu}-\frac{1}{4} {F'}^{\mu\nu}{F'}_{\mu\nu} + \frac{1}{2} m_{A'}^2 {A'}^2 - g_X {A'}_{\mu} \bar{X} \gamma^{\mu} X + \sum_f q_f e (A_{\mu} + \epsilon{A'}_{\mu}) \bar{f} \gamma^{\mu} f,
\end{eqnarray}
in which $m_{A'}$ is the dark photon mass, $g_X$ is the dark U(1) coupling, $q_f$ is the SM fermions electric charge, and $\epsilon$ is the kinetic mixing parameter. The rate for dark photon decays to the SM fermions is
\begin{eqnarray}
  \Gamma(A' \to f\bar{f}) = \frac{\epsilon^2 q_f^2 \alpha (m_{A'}^2 + 2m_f^2)}{3m_{A'}} \sqrt{1-\frac{4m_f^2}{m_{A'}^2}},
  \label{eq:dphot:dec_rate}
\end{eqnarray}
in which $\alpha$ is the electromagnetic structure constant. We will use Eq.~\eqref{eq:dphot:dec_rate} to translate the decay probability $P_{\rm dec}$ in section~\ref{sec:generic} in terms of $(m_{A'}, \epsilon)$. As shown in~\cite{Feng:2016ijc}, relic abundance of the DM ($X$) can be used to fix the dark charge $g_X$, in terms of the rest of the parameters. We will assume the freeze-out scenario which yields $\alpha_X = g_X^2 / (4\pi)= 0.035 (m_X / {\rm TeV})$, while $\epsilon$ and $m_{A'}$ are constrained by the signals from the Sun. For a given set of parameters $(m_X, m_{A'}, \epsilon)$, we follow the procedure in~\cite{Feng:2016ijc} and find the annihilation rate of the DM in the Sun ($\Gamma_{\rm ann}$). We then use the limits from section~\ref{sec:generic} to plot the corresponding contours in $(m_{A'}, \epsilon)$. In calculating the capture rate, we note that the contact interaction limit fails for lighter dark photons ($m_{A'} \sim \mathcal{O} (\textrm{MeV})$). Therefore, we use the following differential cross section for a nucleus ($N$) with mass $m_N$, atomic number $Z_N$, and atomic mass $A_N$
\begin{equation}
    \frac{d\sigma_N}{d E_R} = 8\pi \varepsilon^2 \alpha_X \alpha Z_N^2 |F_N|^2 \frac{m_N}{w^2\left(2m_N E_R + m_{A'}^2\right)^2},
\end{equation}
in which $E_R$ is the recoil energy, $w$ is the incoming DM velocity, and $F_N$ is the Helm factor which is given by
\begin{equation}
    |F_N|^2 = \exp \left(-\frac{E_R}{0.114\, {\rm GeV}} A_N^{5/3}\right).
\end{equation}

\begin{figure}
  \centering
 \includegraphics[width=1\textwidth]{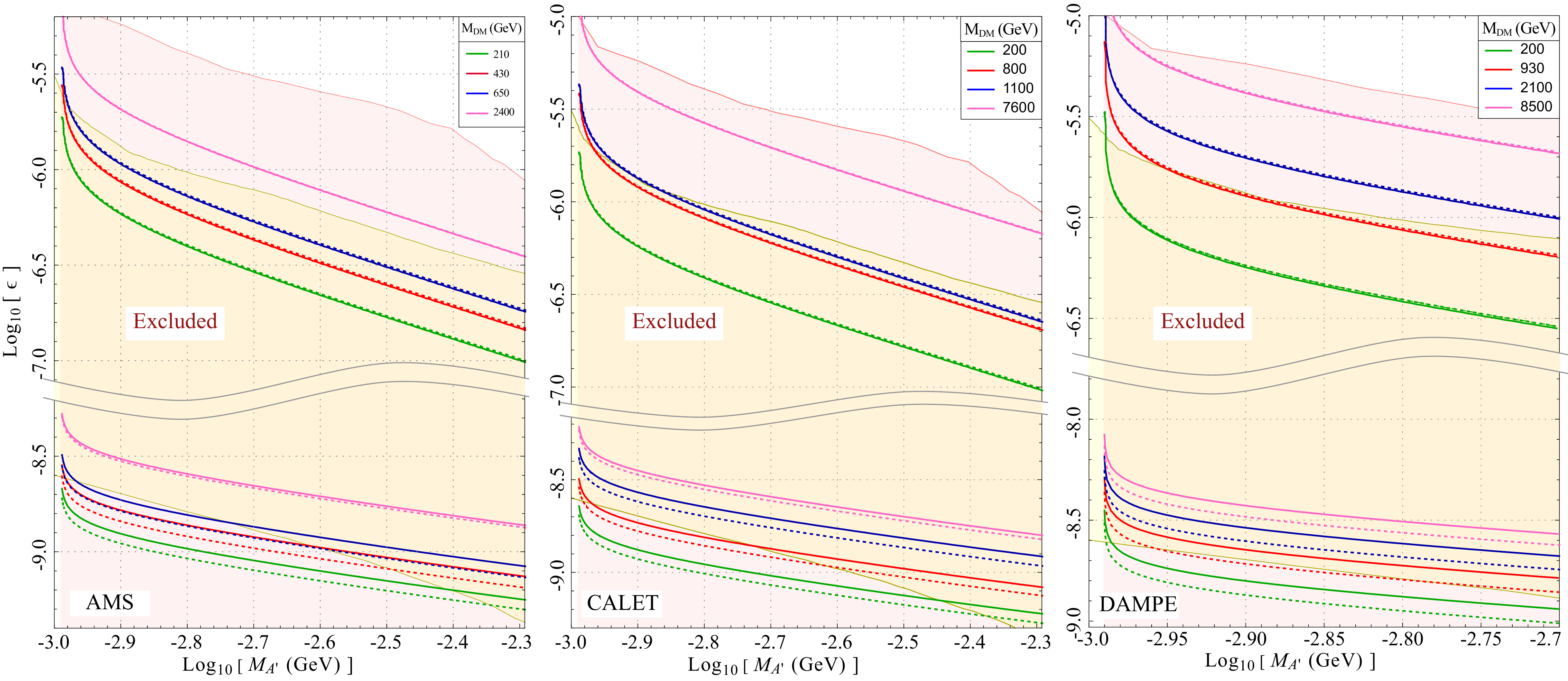}
 \caption{\label{fig:dar:thr:ma}Limits on the kinetic mixing ($\epsilon$) at $95\%$~\acro{C.L.} as a function of the dark photon mass using the current (solid) data and future projections (dashed) in the threshold scenario. Yellow shaded region is the SN1987A cooling bound~\cite{Chang:2016ntp}, and the pink region is from SN1987A explosion bound~\cite{Sung:2019xie}. The areas between the curves are excluded, and the middle part of the plot is removed.}
\end{figure}

\begin{figure}
  \centering
 \includegraphics[width=1\textwidth]{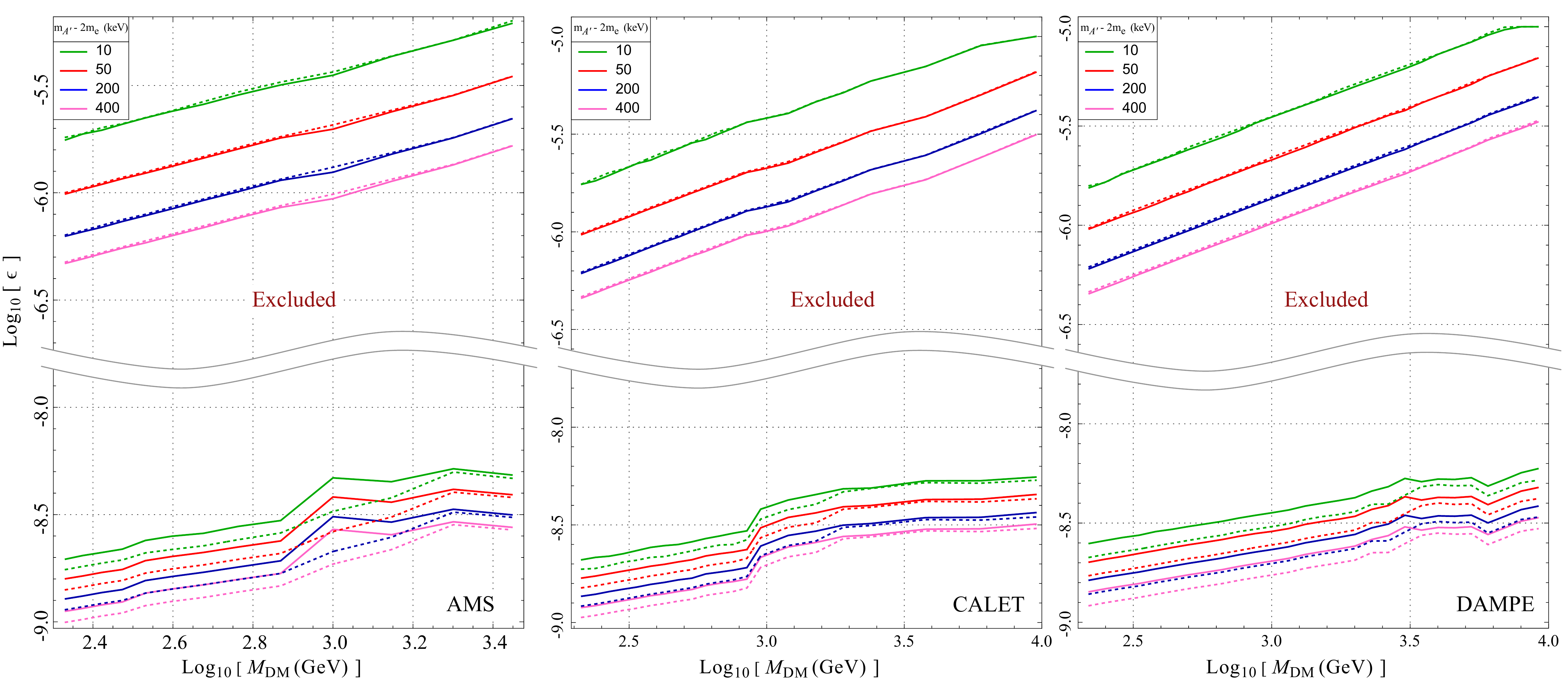}
 \caption{\label{fig:dar:thr:mdm}Limits on the kinetic mixing ($\epsilon$) at $95\%$~\acro{C.L.} as a function of the DM mass using the current (solid) data and future projections (dashed) in the threshold scenario. The areas between the curves are excluded, and the middle part of the plot is removed.}
\end{figure}

\begin{figure}
  \centering
 \includegraphics[width=1\textwidth]{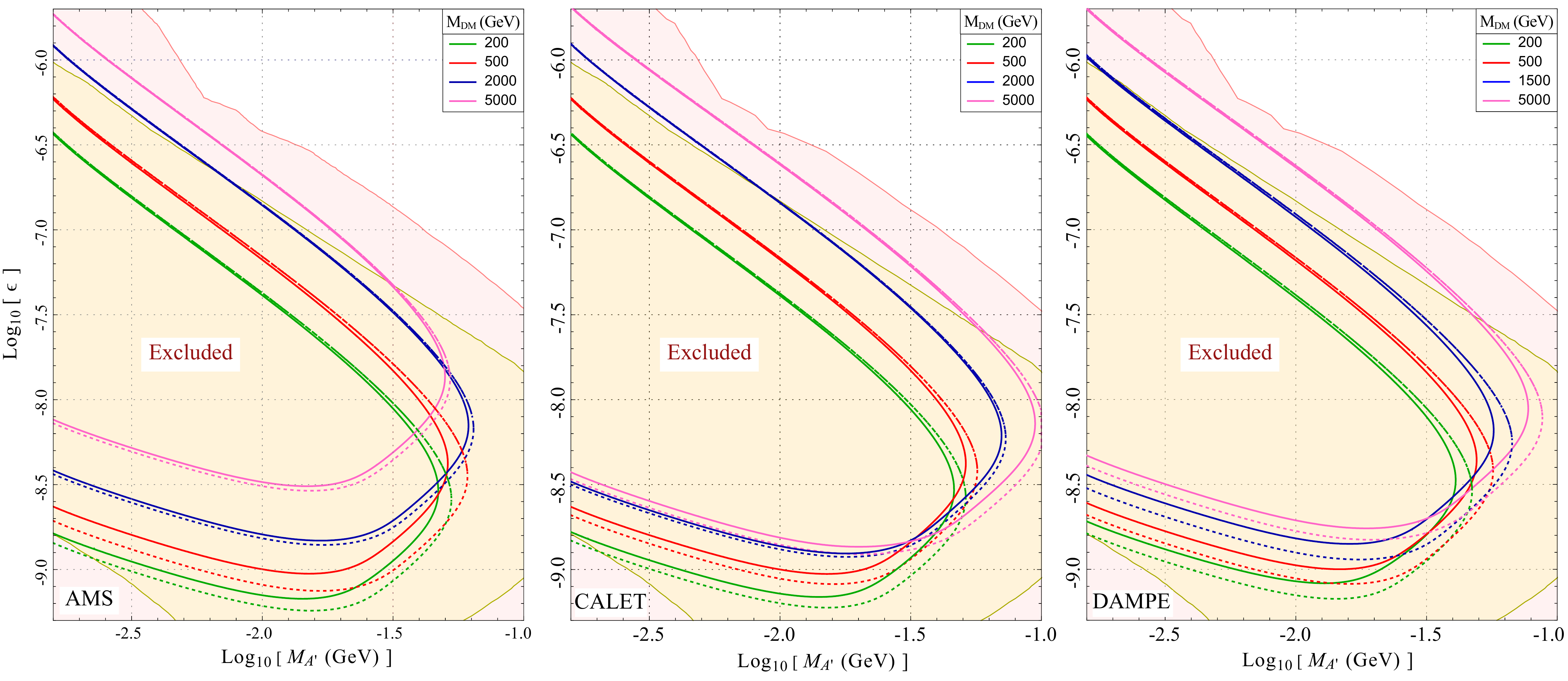}
 \caption{\label{fig:dar:boo:ma}Limits on the kinetic mixing ($\epsilon$) at $95\%$~\acro{C.L.} as a function of the dark photon using the current (solid) data and future projections (dashed) in the boosted scenario. Yellow shaded region is the SN1987A cooling bound~\cite{Chang:2016ntp}, and the pink region is from SN1987A explosion bound~\cite{Sung:2019xie}.}
\end{figure}

\begin{figure}
  \centering
 \includegraphics[width=1\textwidth]{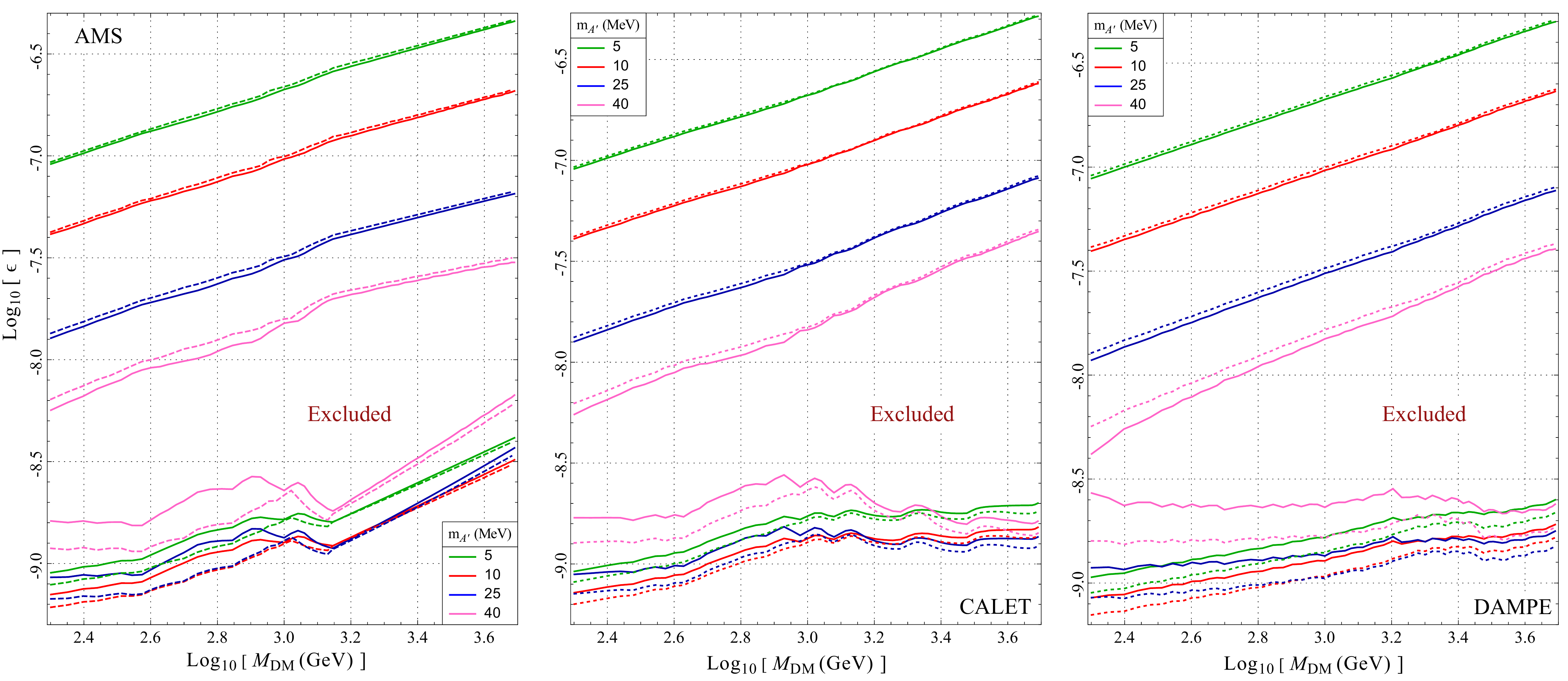}
 \caption{\label{fig:dar:boo:mdm}Limits on the kinetic mixing ($\epsilon$) at $95\%$~\acro{C.L.} as a function of the DM mass using the current (solid) data and future projections (dashed) in the boosted scenario. The areas between the curves are excluded.}
\end{figure}

We have also included the relevant constraints from other experiments for comparison, and removed a horizontal section ($-8 \lessapprox \log(\epsilon) \lessapprox -7 $) of the threshold plots so that the curves are distinguished easier. The constraints on $\epsilon$ as a function of dark matter mass ($M_{\rm DM}$) are plotted in Fig.~\ref{fig:dar:thr:mdm} for the threshold and in Fig.~\ref{fig:dar:boo:mdm} for the boosted case.
The most interesting of these bounds are the ones related to supernova limits. From the results in Fig.~\ref{fig:dar:thr:ma} and Fig.~\ref{fig:dar:boo:ma} we see that our exclusion contours lie within the excluded region from SN1987A explosion bounds~\cite{Sung:2019xie}, and have a lot of overlap with the SN1987A cooling bound~\cite{Chang:2016ntp}. This interesting coincidence can also be observed in the results of Ref.~\cite{Feng:2016ijc}. We note that our signal contours correspond to the $95\%$ exclusion limits, and we have included the updated SN1987A constraints~\cite{Sung:2019xie, Chang:2016ntp}. The similarity between the shape of SN1987A and our excluded region can be explained as follows. The supernova cooling region is limited from below and above because:
\begin{itemize}
  \item Decreasing $\epsilon$ below a certain value will lower its production rate (and energy loss) which loosens the constraint.
  \item Increasing $\epsilon$ beyond a certain value will cause the dark photon decay inside the supernova (no energy escapes) and is unconstrained.
\end{itemize}
On the other hand DM solar signal region is limited from below and above because:
\begin{itemize}
  \item Decreasing $\epsilon$ below a certain value will increase its decay length to values greater than $1$ AU. As a result, most of the mediators decay to $e^{\pm}$ beyond the position of the Earth.
  \item Increasing $\epsilon$ beyond a certain value will cause the dark photon decay into $e^{\pm}$ inside the Sun. In this case the products won't be able to leave the Sun.
\end{itemize}
The above arguments provide a reason for having excluded regions that are bounded from above and below. However, it doesn't specify the boundary values of $\epsilon$ and why they are so close to each other.

\section{Conclusion}
\label{sec:conclusions}
In this work we have analyzed indirect DM signals coming from the Sun in three different space-based detectors. Using the data from cosmic ray detectors in AMS-02, DAMPE and CALET we found the model-independent constraints on the signal strength assuming that no time-dependence has been observed. We then mapped those bounds onto the parameter spaces of generic and dark photon models. We calculated the orbits of each of these detectors, and evaluated their exposure to the Sun as a function of time. We have assumed that if a time-binning of the observed events is done, no time-varying signal is observed, i.e., each time-bin has the same number of events. We then used this non-observation of the signal to set limits on the annihilation rate of the DM in the Sun, and the decay length of the mediator. We have also presented our constraints in terms of spin-dependent DM-proton cross section, and showed that they are better than the limits obtained from direct detection experiments PandaX-II, XENON-1T, and PICO-60. The resulting contours were also translated onto the dark photon parameter space, i.e., $(\epsilon, m_{A'})$ plane. In the dark photon case, our limits lie within the supernova bounds. The generic model results can be used for other models using a similar mapping as $(\Gamma_{\rm ann}, L_{\rm dec}) \longrightarrow (p_1, p_2, \ldots)$, in which $p_1, p_2, \ldots$ are the relevant model parameters. The program used for this work is available on GitHub~\cite{DMSS} and can be modified for studying other space-based detectors and models.

\section*{Acknowledgements}
This work is supported in part by the National Key R\&D Program of China No.~2017YFA0402204, the Key Research Program of the Chinese Academy of Sciences (CAS), Grant NO. XDPB15, the CAS Project for Young Scientists in Basic Research YSBR-006, and the National Natural Science Foundation of China (NSFC) No.~11825506,  No.~11821505, and  No.~12047503. We thank F.~Tanedo for useful comments and discussions. We also thank Y.C.~Ding for his assistance with technology. M. Z.~is supported by the CAS PIFI grant 2019PM0110, and the U.S. Department of Energy, Office of Science, Office of Nuclear Physics, under Award Number DE-FG02-96ER40989.

\appendix

\section{Position of the Sun}
\label{app:sunpos}
In this section we calculate the position of the Sun~\cite{SunTomassetti} as a function of time. We can use the Julian centuries from the epoch J2000 defined as
\begin{align}
  T = \frac{\rm JD - 2451545}{36525},
\end{align}
in which JD is the Julian Day which is the number of days (including fractions) from the beginning of the year $-4712$, and can be computed from a Gregorian date (DD/MM/YY) by
\begin{align}
  {\rm JD } = \floor*{365.25({\rm YY} + 4716)}
  + \floor*{30.6001({\rm MM} + 1)} + {\rm DD} + {\rm B} - 1524.5,
\end{align}
in which if MM $= 1, 2$ then YY should be replaced by YY $ - 1$, and MM by MM $+ 12$, $A = \floor*{\rm YY /100}$, and  $B = 2 - A + \floor*{A/4}$. Then the geometric mean longitude $L_0$ and the mean anomaly $M$ (in degree) are given as
\begin{align}
  L_0 &= 280.4665 + 36000.76983\, T + 0.0003032\, T^2, \\
  M   &= 357.52911 + 35999.05029\, T - 0.0001537\, T^2.
\end{align}
The eccentricity of the Earth’s orbit $e$ is also given by
\begin{align}
  e &= 0.016708634 - 0.000042037\, T - 0.0000001267\, T^2.
\end{align}
The Sun's equation of the center $C$ is given as
\begin{align}
  C =& (1.914602 - 0.004817\, T - 0.000014\, T^2 )\, \sin(M), \\
      &+ (0.019993 - 0.000101\, T )\, \sin(2M) + 0.000289\, \sin(3M),
\end{align}
with all the numbers in units of degree, except the $\sin$ arguments which are in radians. The Sun's geometric longitude is $ \Sun = L_0 + C$, and its true anomaly is $\nu = M + C$. The distance between the centers of the Sun and Earth in AU is given by
\begin{align}
  R = |\vec{R}_{\Sun \to \Earth}| = \frac{1.000001018(1 - e^2)}{1 + e\, \cos(\nu)}.
\end{align}
The Sun’s apparent longitude in ecliptic coordinates $\lambda$ is given by
\begin{align}
  \Omega  =& 125.04452 - 1934.136261\, T, \\
  \lambda =& \Sun - 0.00569 - 0.00478\, \sin (\Omega).
\end{align}
The obliquity of the ecliptic $\epsilon$, i.e., the angle between the ecliptic and the celestial equator can be approximated by
\begin{align}
  \epsilon \approx 23^{\degree}26' 21''.448 - 46''.8150\, T - 0''.00059\, T^2  + 0''.001813\, T^3. 
\end{align}
We can then switch to the equatorial coordinate system in which the right ascension ($\alpha$) and declination ($\delta$) are given as
\begin{align}
  \tan(\alpha) =& \frac{\cos(\alpha)\, \sin(\lambda)}{\cos(\lambda)},\\
  \sin(\delta) =& \sin(\epsilon)\, \sin(\lambda).
\end{align}
The coordinates of Sun in the Geocentric Equatorial Inertial (``GEI") reference frame (Fig.~\ref{fig:GEI}) in units of AU is then given by
\begin{align}
  \vec{R}_{\Sun}(t) = R \times \{\, 
    \cos(\alpha)\, \cos(\delta),\,
    \sin(\alpha)\, \cos(\delta),\, 
    \sin(\delta)\,\}.
\end{align}

\section{Exposure to the Sun}
\label{app:sunexp}
\begin{figure}
  \centering
  \includegraphics[width=0.75\columnwidth]{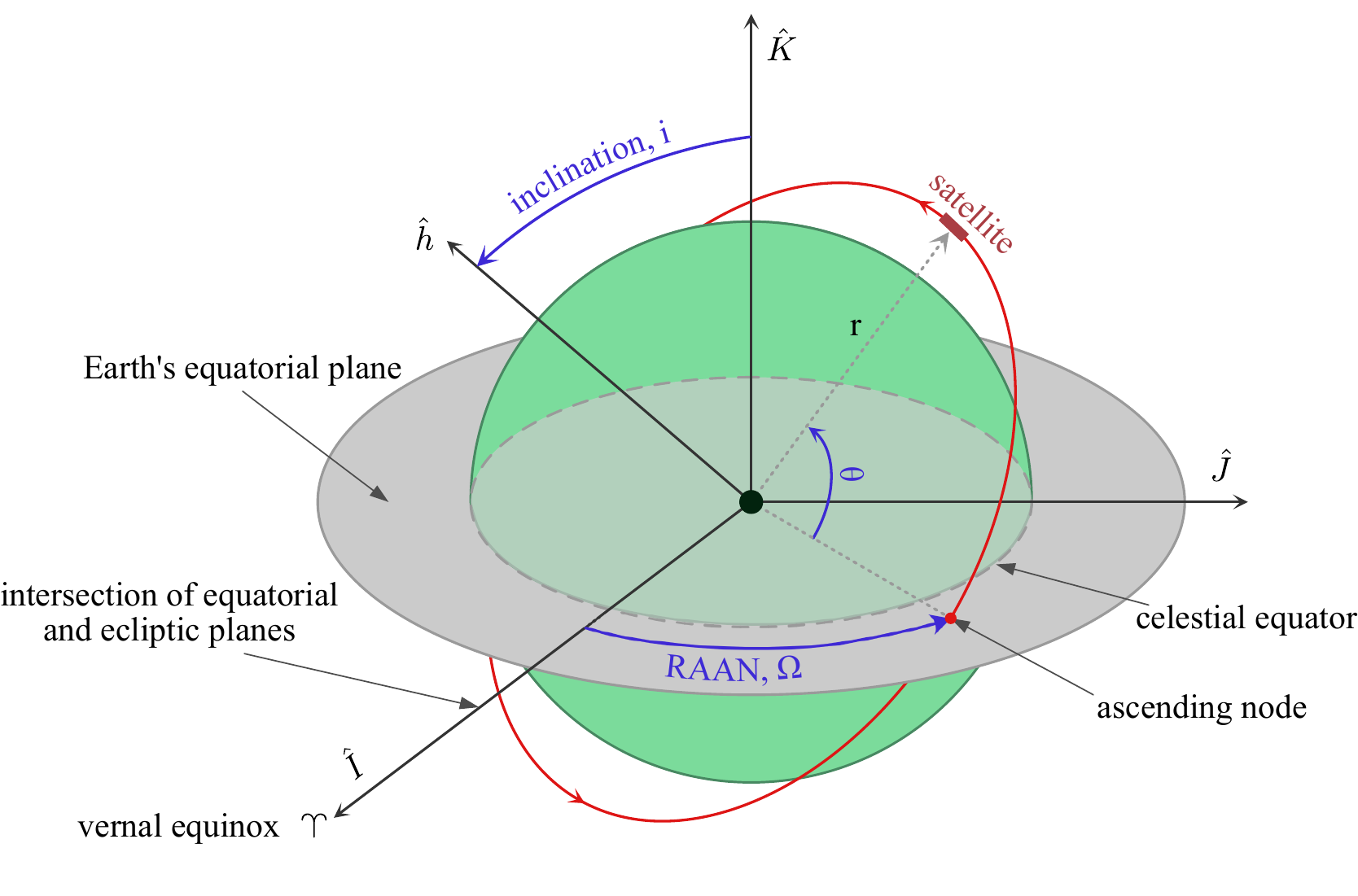}
  \caption{The circular orbit of a detector in geocentric equatorial inertial(``GEI'') reference frame which is described by four parameters: radius ($r$), right ascension of the ascending node ($\Omega$), true anomaly ($\theta$), inclination ($i$). Here, $\hat{h}$ is the normal to the orbit.}
  \label{fig:GEI}
\end{figure}

In this section we focus on finding the exposure of detectors to the Sun ($\xi$) as a function of time. We can write $i(t)$ as
\begin{eqnarray}
  \label{eq:i:angle}
  i(t) = \cos^{-1} \left( \hat{R}_{\Sun \to \rm Sat}(t)\times\hat{n} \right),
\end{eqnarray}
in which $\vec{R}_{\Sun \to \rm Sat}(t) = \vec{R}_{\Sun \to \Earth}(t) + \vec{R}_{\Earth \to \rm Sat}$. Therefore, we first need to find these three vectors $\vec{R}_{\Sun \to \Earth}(t)$, $\vec{R}_{\Earth \to \rm Sat}$, and $\hat{n}$. We use the geocentric equatorial inertial (``GEI") reference frame (Fig.~\ref{fig:GEI}) to represent these vectors. $\vec{R}_{\Sun \to \Earth}(t)$ is just the position of the Sun in GEI frame, i.e., $-\vec{R}_{\Sun}(t)$ which is calculated in the appendix~\ref{app:sunpos}. 

Given the negligible eccentricity of the detector's orbit, we assume that they are circular. $\vec{R}_{\Earth \to \rm Sat}$ which is the position of the detector in GEI frame is then given by
\begin{eqnarray}
  \vec{R}_{\Earth \to \rm Sat}(t) = (r_{\Earth} + h) \times\, \mathbb{R}_{\hat{K}}(\Omega) \times \mathbb{R}_{\hat{I}}(i) \times  \mathbb{R}_{\hat{K}}(\theta) \times \hat{I},
\end{eqnarray}
in which $\theta$ is the true anomaly, $i$ is the inclination, $\Omega$ is the right ascension of the ascending node (RAAN) of the orbit, $\mathbb{R}_{\hat{K}, \hat{I}}$ are rotation matrices along $\hat{K}$ \& $\hat{I}$ directions respectively (Fig.~\ref{fig:GEI}), and $h$ is the mean height of the orbit.

\begin{figure}
  \centering
  \includegraphics[width=0.45\columnwidth]{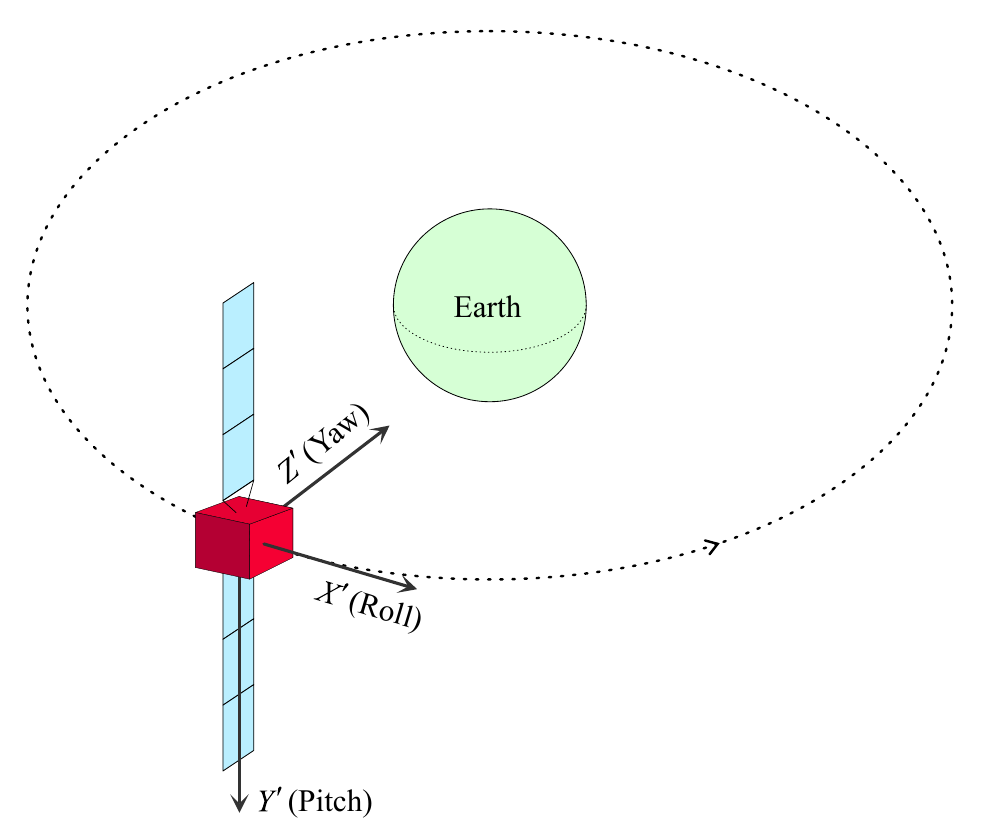}
  \caption{The local detector frame with the definitions of yaw ($Y$), pitch ($P$), and roll ($R$).}
  \label{fig:ypr}
\end{figure}

The local detector frame (see Fig.~\ref{fig:ypr}) is defined by the $Z'$ axis pointing to the Earth (nadir) direction, $X'$ axis being in the direction of motion along the orbit, and $Y'$ as a result of the right-hand-rule. Therefore, we can write these axes in the GEI coordinates as
\begin{eqnarray}
  \hat{X}' =& \mathbb{R}_{\hat{K}}(\Omega) \times \mathbb{R}_{\hat{I}}(i) \times  \mathbb{R}_{\hat{K}}(\theta) \times \hat{J},\\
  \hat{Y}' =& \hat{Z}' \times \hat{X}',\\
  \hat{Z}' =& -\hat{R}_{\Earth \to \rm Sat}(t).
\end{eqnarray}
Consequently, a rotation from the local frame to the GEI coordinates can be defined as
\begin{eqnarray}
  \mathbb{R}'_{i,j} = \langle \hat{e}_i | \hat{e}'_j\rangle, 
\end{eqnarray}
with $\hat{e} = {\hat{I}, \hat{J}, \hat{K}}$ and $\hat{e}' = {\hat{X}', \hat{Y}', \hat{Z}'}$. The normal vector to the surface of the detector in the local detector frame (Fig.~\ref{fig:ypr}) is also given by 
\begin{eqnarray}
  \hat{n}' = \mathbb{R}_{\rm YPR}\times (0, 0, -1),
\end{eqnarray}
in which $\mathbb{R}_{\rm YPR}$ is a rotation defined by the local yaw ($Y$), pitch ($P$), roll ($R$) coordinates of the detector with respect to the orbit. We can finally express $\hat{n}'$ in the GEI coordinates as $\hat{n} = \mathbb{R}'\times \hat{n}'$. We can now use Eq.~\ref{eq:exp} and \ref{eq:i:angle} to evaluate the exposure.

The field-of-view (FOV) condition in the case of a conical approximation, can be written as $\pi - i(t) < \theta_{\rm FOV}$. We also include the effects of the Earth's shadow by excluding the times when the detector is in the Earth's shadow. There are two condition that must be satisfied for the detector to be in the Earth's shadow: (a)  it should be on the night-side of the Earth, (b) it should be inside the cylinder centered at the Earth with radius $r_{\Earth}$, and axis pointing to the Sun, i.e., 
\begin{eqnarray}
  {\rm (a)} & \vec{R}_{\Sun}\times \vec{R}_{\rm Sat} < 0,\\
  {\rm (b)} & \sin(\arccos(\hat{R}_{\Sun}\times \hat{R}_{\rm Sat}))\times |R_{\rm Sat}| < r_{\Earth}.
\end{eqnarray}
This concludes the evaluation of the exposure to the Sun. In what follows, we will examine specific experiments, and evaluate their exposure to the Sun.

\subsection{Alpha Magnetic Spectrometer-02 (AMS-02)}
\label{app:sunexp:ams}
For AMS-02 exposure time interval, we use the results from May 19, 2011 to November 12, 2017~\cite{Aguilar:2019ksn}, and for the orbital situation of the AMS-02 on ISS we use Ref.~\cite{XIE201650}. ISS orbit inclination is $51.6^{\degree}$, the orbit height is about $400$ km, and its period is $92.68$ minutes. The orbits are assumed to be circular as mentioned before. ISS is situated at $(Y, P, R) = (-4^{\degree}, -2^{\degree}, +0.7^{\degree})$, relative to the orbit and AMS-02 is tilted with an extra $-12^{\degree}$ roll relative to the ISS~\cite{XIE201650}. The FOV for AMS-02 is about $48^{\degree}$ half cone angle~\cite{ams:report}. The effective detector acceptance is given by $A_{\rm eff} = A_{\rm geom}\epsilon_{\rm sel}(1+\delta)$~\cite{Aguilar:2014fea}, in which $A_{\rm geom} \approx 550\, {\rm cm}^2 {\rm sr}$ is the geometric acceptance, $\epsilon_{\rm sel}$ is the event selection efficiency which we approximate by a linear fit to the values given in Ref.~\cite{Aguilar:2014fea}:

\[ \epsilon_{\rm sel} = \begin{cases} 
   1& \qquad\qquad \quad \quad \,\,\,\,\,  E\leq 10 \,\,\,{\rm GeV} \\
  (0.908 - 0.00078\times E)& \qquad 10 \,\,\,{\rm GeV} \leq E\leq 100 \,{\rm GeV} \\
  (0.844 - 0.00014\times E)& \qquad 100 \,{\rm GeV} \leq E 
\end{cases}
\]
$\delta$ is a correction factor to the acceptance for which we assume $\delta \approx -0.035$. We also assume a linear behavior given the values in Ref.~\cite{Aguilar:2018ons}         

\[ r_t = \begin{cases} 
  0& \qquad\qquad \quad \quad \,\,\,\,\,  E\leq 1 \,\,\,{\rm GeV} \\
  0.06& \qquad \,\,1 \,\,\,\,{\rm GeV} \leq E\leq 2 \,\,\,{\rm GeV} \\
  (2.242\times E + 1.515)\times 0.01& \qquad \,\,2\,\,\, \,{\rm GeV} \leq E \leq 35 \, {\rm GeV}\\
  0.8& \qquad \,\,35 \,{\rm GeV} \leq E
\end{cases}
\]
in which $T = r_t\times\Delta T$, $\Delta T$ is the total time interval ($1$ Bartel in Ref.~\cite{Aguilar:2018ons}), and $T$ is the effective exposure time. Note that these values are assuming one Bartel ($27$ days) time intervals. However, we assume the same expression for similar ranges of time intervals. The errors arising here can also be compensated in form of an overall correction factor at the end of the analysis.

\begin{figure}[htb]
  \centering
 \includegraphics[width=1\textwidth]{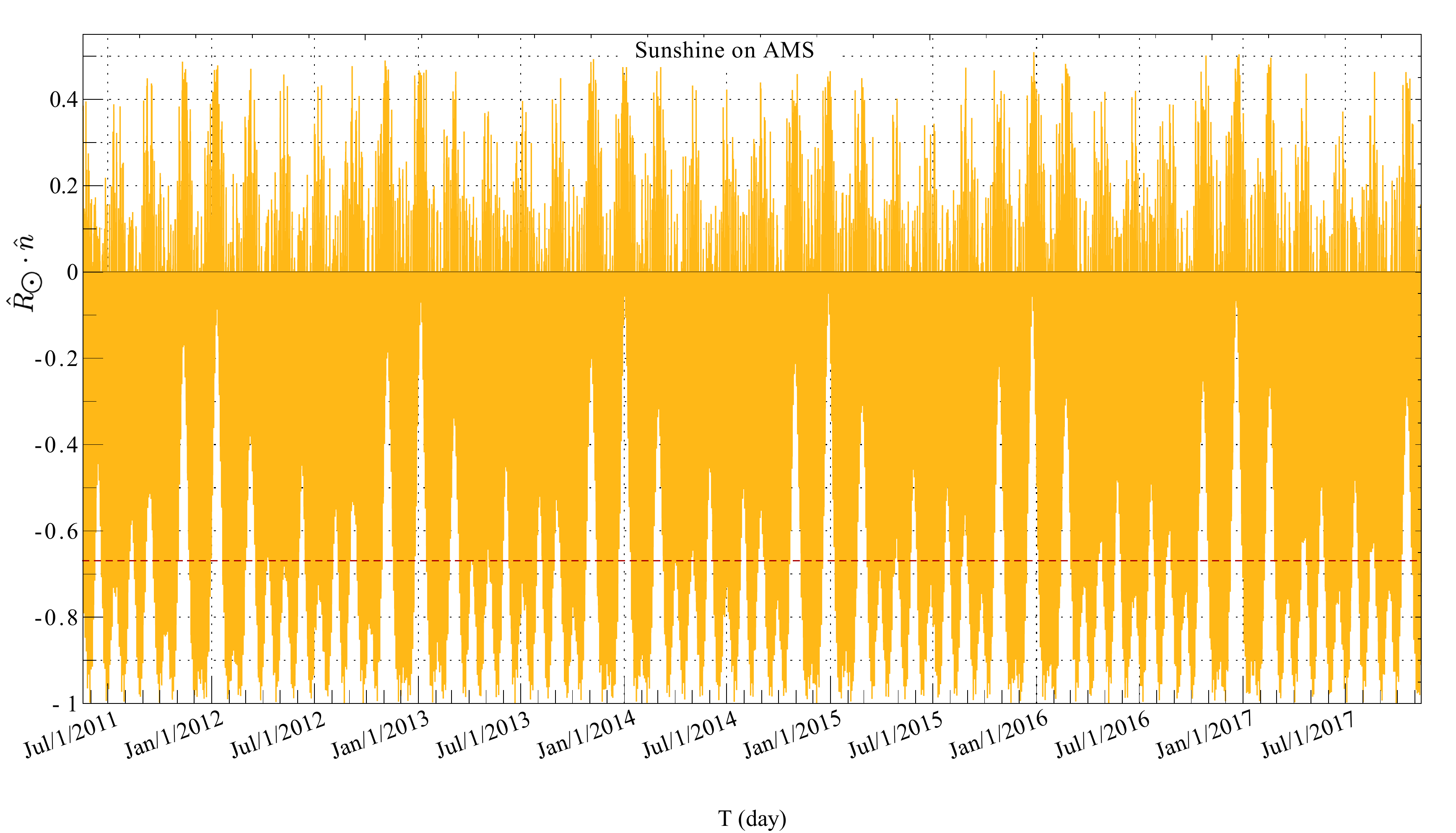}
 \caption{\label{fig:ams:sunshine} The sunshine ($ \hat{n}\cdot \vec{R}_{\rm Sun \to \rm Sat}$) on the AMS-02 detector between May 19, 2011 to November 13, 2017. The red-dashed line corresponds to $\theta_{\rm FOV} = 48^\degree$ from the normal vector ($\hat{n}$).}
\end{figure}

\begin{figure}[htb]
  \centering
 \includegraphics[width=1\textwidth]{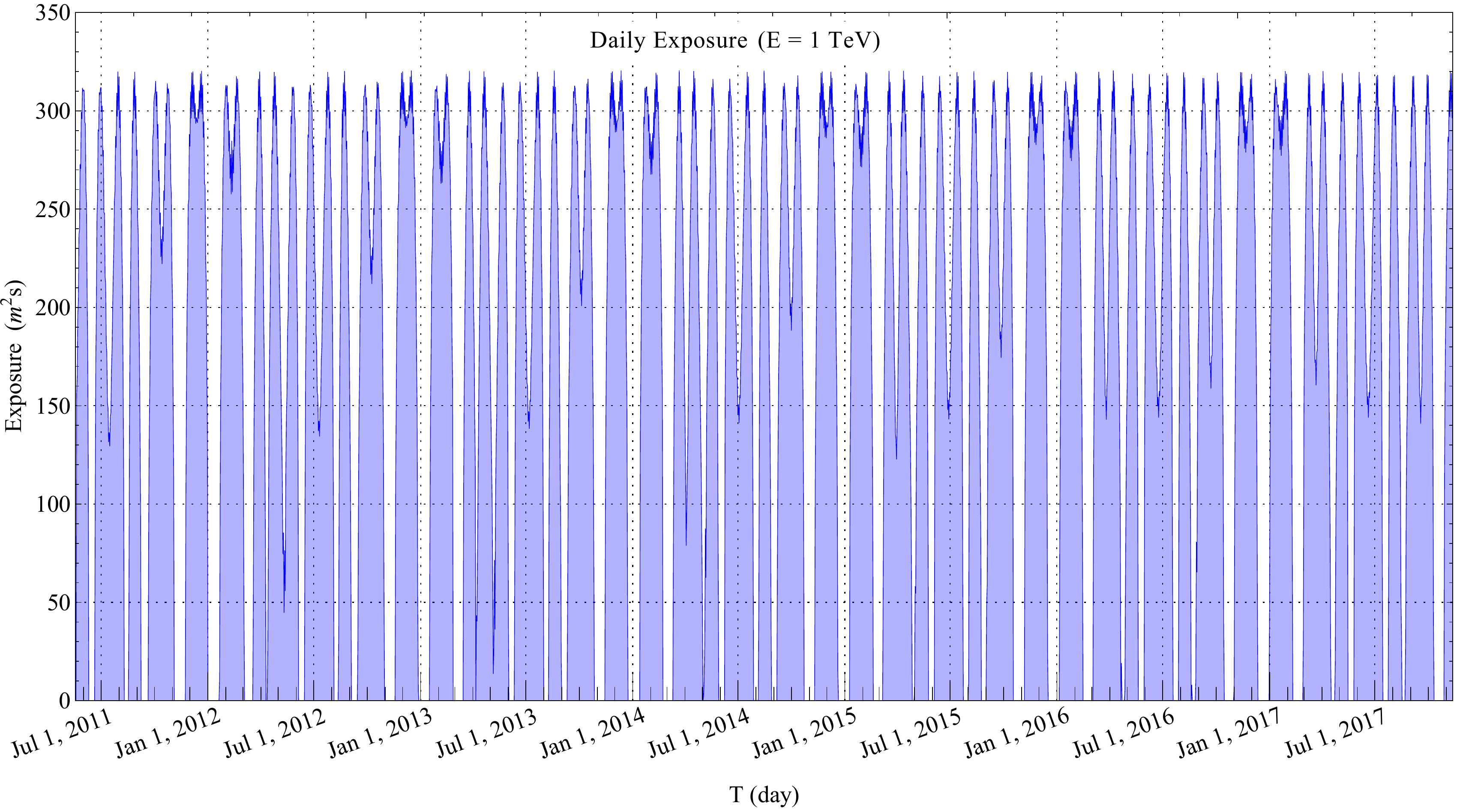}
 \caption{\label{fig:ams:exp}The AMS-02's daily exposure ($\xi$) to the Sun at $E = 1$ TeV.}
\end{figure}

The variation of cosine of the incident angle of rays from the Sun on the AMS detectors, i.e., ``sunshine'' ($\hat{n}\cdot \vec{R}_{\rm Sun \to \rm Sat}$) is plotted in Fig.~\ref{fig:ams:sunshine}, and the AMS exposure to the Sun ($\xi$) is shown in Fig.~\ref{fig:ams:exp}.
In comparison with the yearly average exposure quoted in Ref.~\cite{Feng:2016ijc}, i.e., $\xi_{\Sun} = 6.3 \times 10^4\, {\rm m}^2 {\rm s}$, we get: $\xi_{\Sun} (1 \,\rm TeV) = \{ 6.573, 6.560, 6.560\} \times 10^{4} \, {\rm m}^2 {\rm s}$ for the years 2012, 2013, and 2014 respectively. We get the same yearly average exposure as in~\cite{Feng:2016ijc} at $E = 1.2$ TeV, i.e., $\xi_{\Sun} (1.2 \,\rm TeV) = 6.3 \times 10^4\, {\rm m}^2 {\rm s}$. 

\subsection{CALorimetric Electron Telescope (CALET)}
\label{app:sunexp:calet}
\begin{figure}[htb]
  \centering
 \includegraphics[width=1\textwidth]{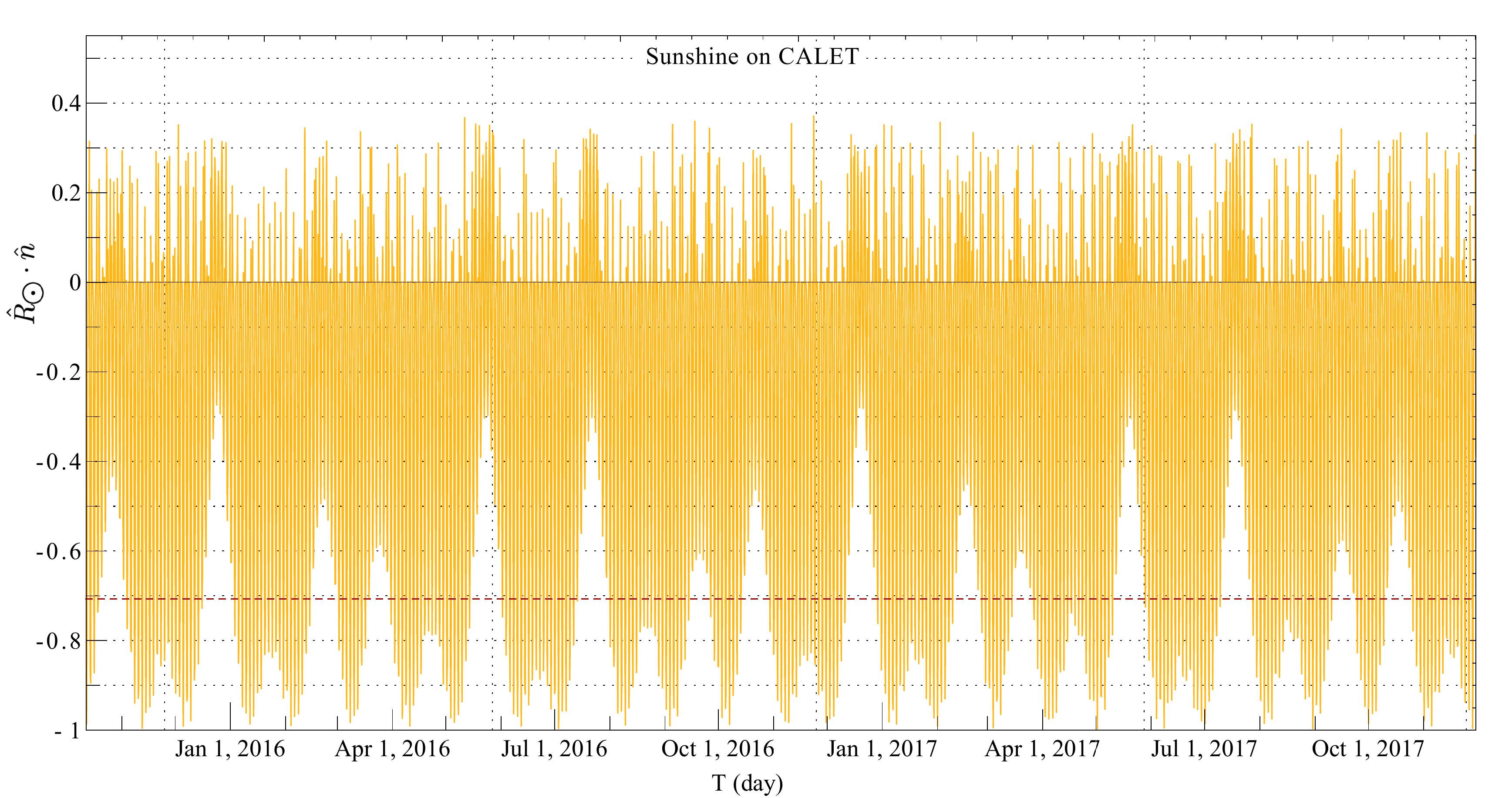}
 \caption{\label{fig:calet:sunshine}The sunshine on the CALET detector between Oct 13, 2015 to Nov 31, 2017. The red-dashed line corresponds to $\theta_{\rm FOV} = 45^\degree$.}
\end{figure}

\begin{figure}[htb]
  \centering
 \includegraphics[width=1\textwidth]{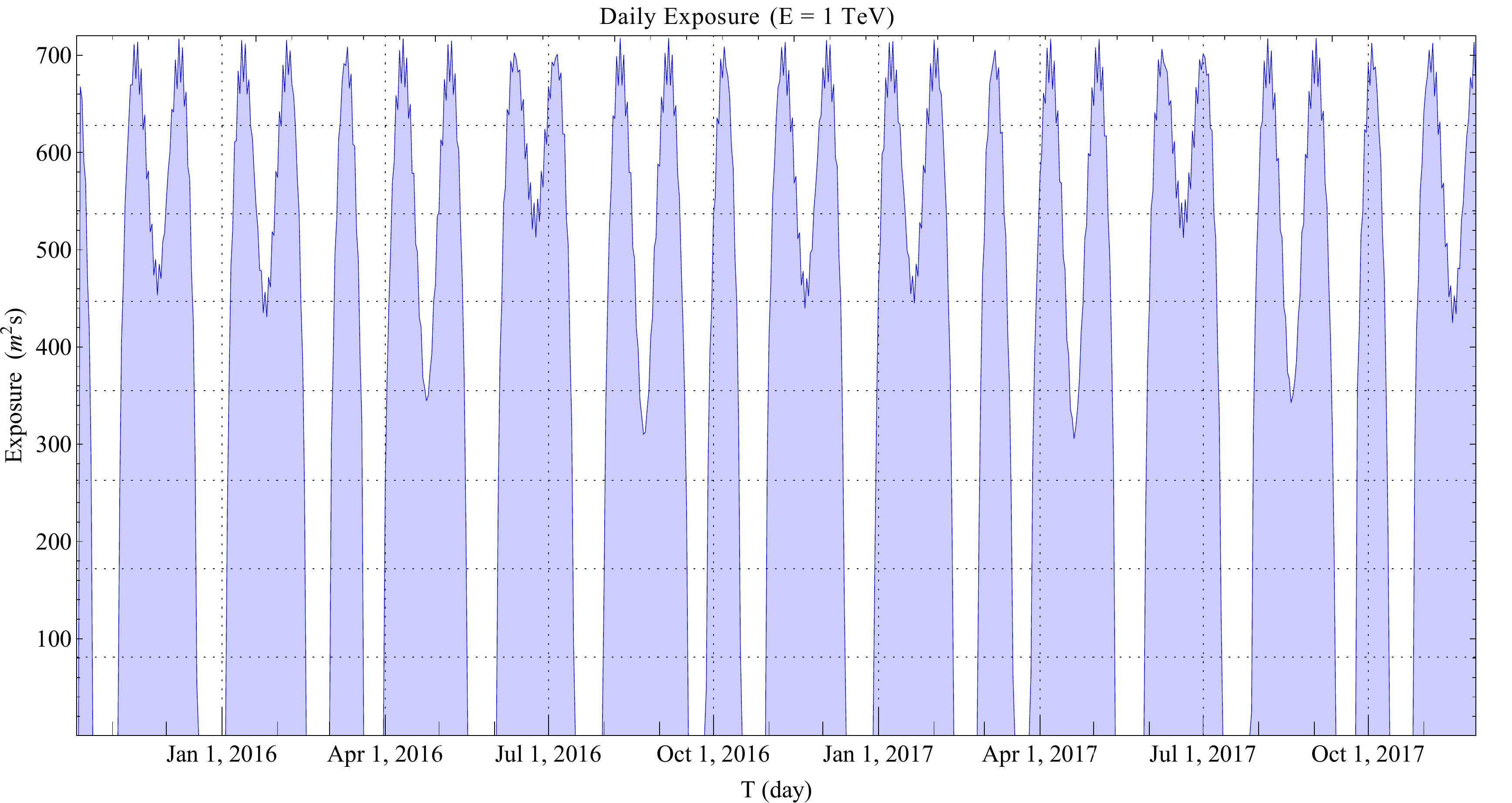}
 \caption{\label{fig:calet:exp}The CALET's daily exposure ($\xi$) to the Sun at $1$ TeV.
 }
\end{figure}

For the CALET, we assume 780 days of exposure from October 13, 2015 to November 30, 2017~\cite{Adriani:2018ktz}. CALET is also on the ISS, but it's not tilted relative to the ISS (like AMS-02). The FOV is about $45^\degree$ from zenith~\cite{Adriani:2018ktz}. The live time fraction for CALET in~\cite{Adriani:2018ktz} is $84\%$ for the whole operation time. Here, we are going to assume the same value for smaller time intervals, which is justified given a continuous and stable data collection~\cite{Adriani:2017efm}. The geometrical acceptance for HE electrons is $A_{\rm geom} \approx 1040 \, {\rm cm}^2 {\rm sr}$~\cite{Adriani:2018ktz}, and it's almost independent of energy ($E > 10$ GeV)~\cite{Adriani:2017efm}. The total detection efficiency is $\epsilon \sim73\%$, and it's very stable for energies up to 3 TeV~\cite{Adriani:2017efm}. Therefore, we will take the effective acceptance of CALET to be $A_{\rm eff} = \epsilon \times A_{\rm geom} \approx 7.592\times 10^{-2}\, {\rm m}^2 {\rm sr}$. The sunshine ($\hat{n}\cdot \vec{R}_{\rm Sun \to \rm Sat}$) on CALET's detectors is plotted in Fig.~\ref{fig:calet:sunshine}, and its exposure to the Sun ($\xi$) is shown in Fig.~\ref{fig:calet:exp}.

\subsection{DArk Matter Particle Explorer (DAMPE)}
\label{app:sunexp:dampe}
\begin{figure}[htb]
  \centering
 \includegraphics[width=1\textwidth]{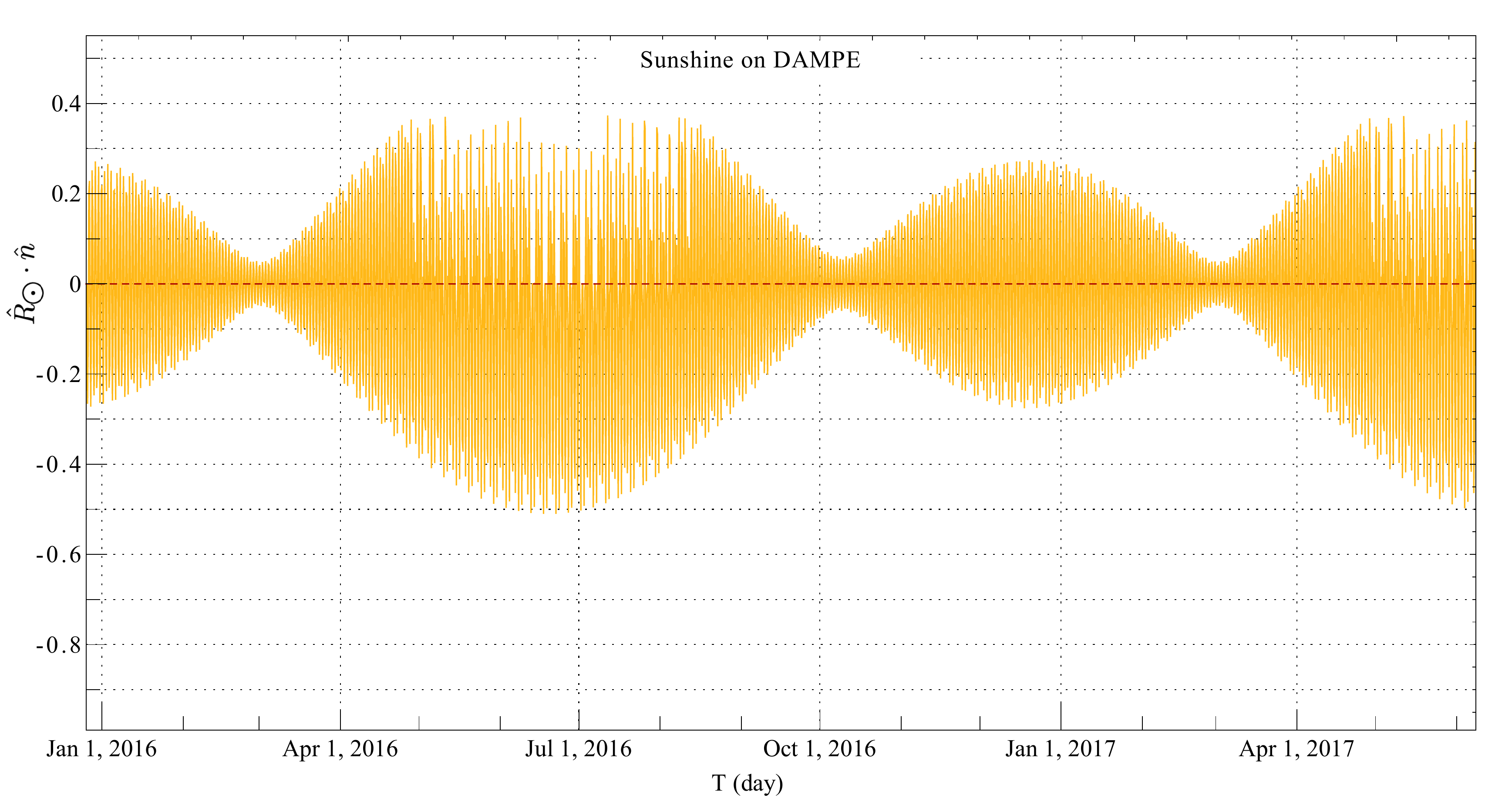}
 \caption{\label{fig:dampe:exp:sunshine}
 The sunshine ($ \hat{n}\cdot \vec{R}_{\rm Sun \to \rm Sat}$) on the DAMPE detector between December 27, 2015 to June 9, 2017. The red-dashed line corresponds to $\theta_{\rm FOV} = 90^\degree$ from the normal vector ($\hat{n}$). 
 }
\end{figure}

\begin{figure}
  \centering
 \includegraphics[width=1\textwidth]{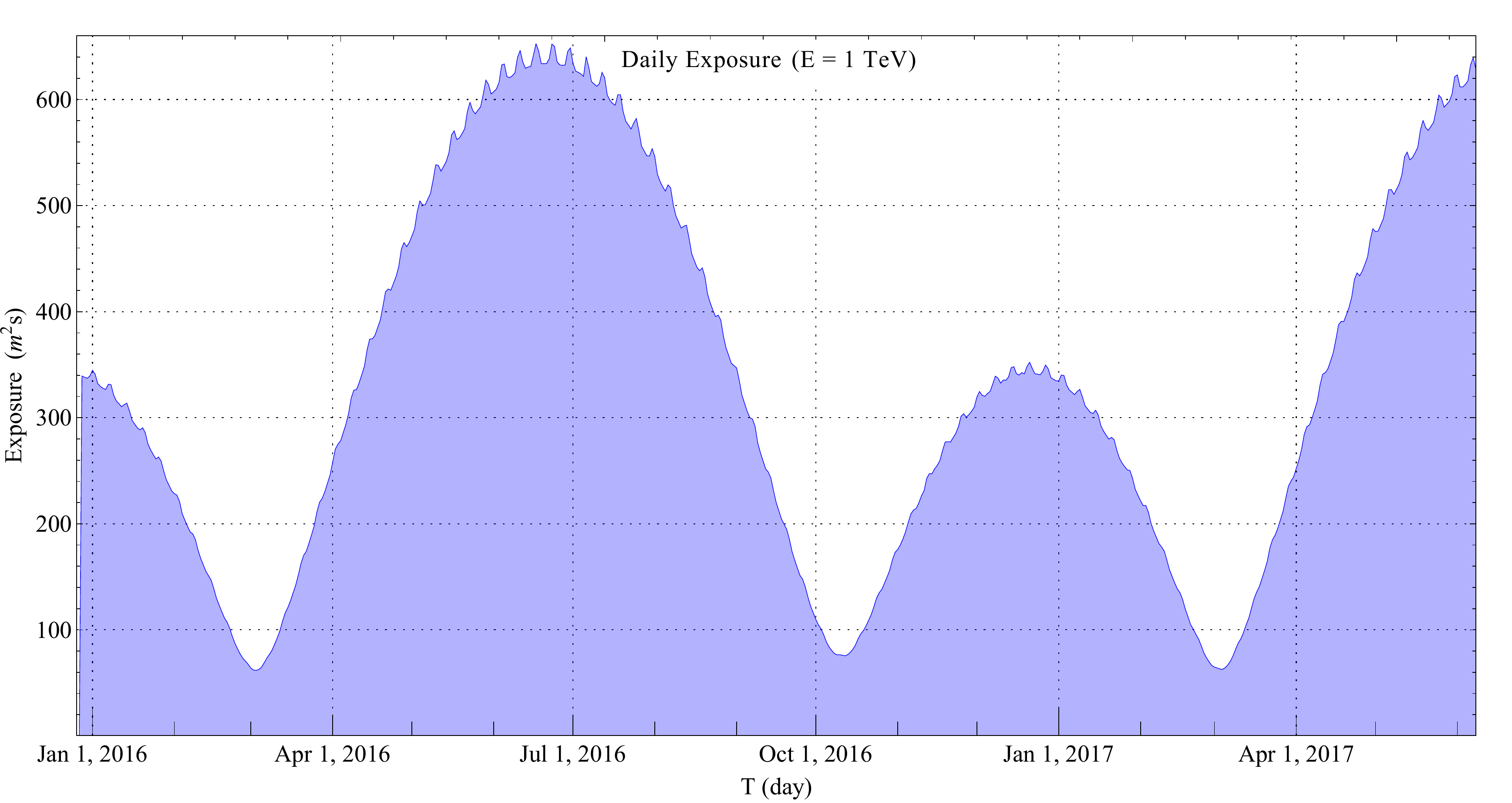}
 \caption{\label{fig:dampe:exp:exp}
 The DAMPE's daily exposure ($\xi$) to the Sun at $E = 1$ TeV.
 }
\end{figure}

DAMPE is on a sun synchronous orbit at $h=500$ km, with an orbit period of $95$ mins. The orbit inclination is $97.406^{\degree}$, with local time descending node of 6:30 AM~\cite{dampe:attitude}. We will assume a $530$ days exposure between December 27, 2015 and June 8, 2017~\cite{Ambrosi:2017wek}. We use the effective acceptance from Table 1 in Ref.~\cite{Ambrosi:2017wek} for each energy bin, which is $\sim0.24\, {\rm m}^2 {\rm sr}$ for $E>1$ TeV. The sunshine ($\hat{n}\cdot \vec{R}_{\rm Sun \to \rm Sat}$) on DAMPE's detectors is plotted in Fig.~\ref{fig:dampe:exp:sunshine}, and its exposure to the Sun ($\xi$) is shown in Fig.~\ref{fig:dampe:exp:exp}.

\clearpage
\bibliographystyle{JHEP}
\bibliography{DMSolSig}

\end{document}